\documentclass[%
 reprint,
 amsmath,amssymb,
 aps,
]{revtex4-1}

\usepackage{graphicx}
\usepackage{dcolumn}
\usepackage{bm}

\begin{document}

\preprint{APS/123-QED}

\title{Ultra Thin Films of Yttrium Iron Garnet with Very Low Damping: A Review}

\author{G. Schmidt}
\thanks{Corresponding author:georg.schmidt@physik.uni-halle.de}
\affiliation{Institut f\"{u}r Physik, Martin-Luther-Universit\"{a}t Halle-Wittenberg, D-06120 Halle, Germany\\}
\affiliation{Interdisziplin\"{a}res Zentrum f\"{u}r Materialwissenschaften, Martin-Luther-Universit\"{a}t Halle-Wittenberg, D-06120 Halle, Germany\\}

\author{C. Hauser}
\affiliation{Institut f\"{u}r Physik, Martin-Luther-Universit\"{a}t Halle-Wittenberg, D-06120 Halle, Germany\\}

\author{Philip Trempler}
\affiliation{Institut f\"{u}r Physik, Martin-Luther-Universit\"{a}t Halle-Wittenberg, D-06120 Halle, Germany\\}
\author{M. Paleschke}
\affiliation{Institut f\"{u}r Physik, Martin-Luther-Universit\"{a}t Halle-Wittenberg, D-06120 Halle, Germany\\}
\author{E. Th. Papaioannou}
\affiliation{Institut f\"{u}r Physik, Martin-Luther-Universit\"{a}t Halle-Wittenberg, D-06120 Halle, Germany\\}


\date{\today}


\begin{abstract}
Thin  Yttrium Iron Garnet (YIG) is a promising material for integrated magnonics. In order to introduce YIG into nanofabrication processes it is necessary to fabricate very thin YIG films with a thickness well below 100 nm while retaining the extraordinary magnetic properties of the material, especially its long magnon lifetime and spin wave propagation length. Here, we give a brief introduction into the topic and we review and discuss the various results published over the last decade in this area. Especially for ultrathin films it turns out that pulsed layer deposition and sputtering are the most promising candidates. In addition, we discuss the application of room temperature deposition and annealing for lift-off based nanopatterning and the properties of nanostructures obtained by this method over the past years.
\end{abstract}

\maketitle   

\section{Introduction}
In the past decade the field of magnonics has become more and more attractive. Using magnons for information processing would allow for low dissipation data processing devices and hold the promise of complex functionality~\cite{Kruglyak2010,Khitun2012,Chumak2015,Lenk2011}. Novel concepts have already been proposed including magnon logic circuits, magnon transistor, reconfigurable magnonic devices, spin-wave frequency filters, signal processing and computing~\cite{Khitun2010,Fischer2017,Krawczyk2014,Chumak2014,Yu2016,Urazhdin2014}.
At the same time the use of ferromagnetic materials immediately brings to mind the possibility of non-volatile data storage or non-volatile programmability. Even the use of magnons in the context of quantum information processing has been attempted~\cite{Tabuchi2015}. All these concepts, however, rely on one prerequisite. They all need to be integrated and in order to compete as a post-CMOS technology, they somehow need to approach the density of ultra large scale integration. While multi functional devices can by some means relax the constraints which nowadays drive lithography for CMOS processes towards 10\,nm resolution, it is still obvious that any post-CMOS technology needs to be able to address at least sub-100\,nm lateral resolution. From the point of view of commercial lithography this is not a problem at all. The challenge, however is on the materials side. Thin film metallic ferromagnets can easily be patterned with the required resolution as has been demonstrated in MRAM processing. Nevertheless, in order to realize complex magnonics, it is necessary to have very low damping and very long spin relaxation lengths in the ferromagnet of choice. Here, the range of $\alpha\leq 5\times 10^{-4}$ is still restricted to the ferrimagnet Yttrium Iron Garnet (YIG). This material is well known and well studied since several decades and as a bulk or thick film material can exhibit Ferromagnetic Resonance (FMR) linewidths of 15\,$\mu$T (full width at half maximum-FWHM) at 9.6\,GHz~\cite{Glass1976} and show a damping of $\alpha < 3\times10^{-5}$. Many concepts of magnonic devices have been realized in YIG, however, mostly in films with a thickness much larger than 100\,nm. The main reason lies in the fabrication technology for thin YIG films. YIG films of several micron thickness have traditionally been deposited using liquid phase epitaxy (LPE). This method still yields the lowest damping and linewidth in non-bulk YIG. However, there seems to be a lower thickness limit for LPE-YIG under which the quality starts to degrade. A large effort in this respect now makes 100\,nm YIG films with damping below $\alpha = 10^{-4}$ commercially available on substrates of 3" diameter or more, but still thinner films with similar quality cannot be reliably fabricated by LPE. This is a large drawback because for nanopatterning the aspect ratio is always a critical factor because for many etching processes it limits the lateral structures size to more than the layer thickness. Large aspect ratio in nanopatterning is mainly achieved by dry etching. No etching technique is known that yields an aspect ratio of more than 1:1 in YIG and still retains its magnetic properties. So one would rather want to use a layer thickness which is well below the lateral target resolution. As a consequence alternative deposition techniques are needed.

A major competing deposition technique is pulsed laser deposition (PLD). This technique uses a target which typically has the stoichiometry of the film that is to be deposited. The material is ablated by a ns laser pulse and transferred in a so-called plasma plume to the substrate which is placed opposite to the target. The target can be heated yielding the necessary surface mobility for the different impinging species to arrange in the desired lattice structure during the deposition. For YIG this method was already demonstrated more than 20 years ago, for example in 1993 by Dorsey \textit{et al.}~\cite{Dorsey1993}. Although for thick films the quality of PLD deposited films is not as high as for LPE grown material, for ultra thin films of well below 100\,nm PLD seems to yield better results. Only recently, further progress was made using room temperature PLD and subsequent annealing, resulting in layers with a damping of $\alpha = (6.15 \pm 1.50) \times 10^{-5}$~\cite{Hauser2016}. Room temperature deposition and annealing also allows to use the more industry compatible process of magnetron sputtering to achieve very high quality~\cite{Chang2014}, which is not possible with sputtering at high temperature.

The remaining part of the paper is structured as follows: First, we will briefly discuss the assessment of magnonic quality and possible quality criteria. Concentrating on film thicknesses below 100\,nm and setting the lower limit of the quality to a damping of $\alpha\leq 4 \times 10^{-4}$  or an FMR linewidth of less than 400\,$\mu$T, or both, we will then discuss results published since 2012. In the sequence of their appearance we will start with PLD at elevated substrate temperature. Then results obtained by off-axis spttering, sputtering at room temperature with subsequent annealing, and finally by PLD at room temperature will be described. Because room temperature deposition processes offer the opportunity for novel nanopatterning strategies, another section is dedicated to lift-off patterning of YIG.

In the end, we present a table with the parameters for all layers presented in the papers discussed here as far as details were given by the respective authors.

\section{Damping, linewidth, and propagation length}
In order to assess the quality of a magnetic film for magnonics, different criteria can be used. Obviously for magnonics, magnons need to propagate. So the propagation (or decay) length $l_{\text{decay}}$ can be an important parameter and a large propagation length is certainly a good argument to use a material.
The decay length is the relevant length scale for the decay of the spin wave amplitude.

\begin{equation}
A(x)=A(0)e^{-\frac{x}{l_\text{decay}}}
\end{equation}

On the other hand, it is not necessary to achieve a propagation length which is far beyond the size of the target device. It is merely necessary for $l_{\text{decay}}$ to be long enough so that at the end of the propagation the signal is sufficiently large for further processing including all possible damping effects which may occur on the propagation path. The propagation length in turn is related to the magnon lifetime $\tau_{\text{magnon}}$ and the group velocity $v_{\text{g}}$ by the equation:

\begin{equation}\label{vg}
l_{\text{decay}}=\tau_{\text{magnon}} v_{\text{g}}
\end{equation}

This is quite important because it imposes much more severe constraints on thin films than on thick films. For small k-vectors the magnon dispersion for thin films is much more flat (Fig.\ref{Dispersion}) than for thick films and the group velocity $\partial\omega/\partial k$ is much lower. As a consequence, for the same propagation length in a thin film much larger magnon lifetimes are needed than in a thick film. The lifetime, however, can be calculated from the Gilbert damping $\alpha$ and/or the resonance linewidth $\mu_{\text 0}\Delta H_{0}$ by the following formula~\cite{Yu2014}:

\begin{equation}\label{lifetime}
\tau _{0}=\frac{1}{2\pi \alpha f_{\text{res}}}
\end{equation}

where f$_{\text{res}}$ is the eigenfrequency of the respective mode. Although thinner films are less favourable in terms of $v_{\text{g}}$ even for thin films propagation lengths have been presented which seem long enough for magnonics applications. In the following a number of examples are given from layers which are also described later in detail. All values are obtained for Damon Eshbach modes. The results strongly differ because the lifetime is inversely proportional to damping and frequency, respectively (eq. \ref{lifetime}). So Talalaevskij \textit{et al.} \cite{Talalaevskij2017} use a 49 nm thick film which due to metal coverage and spin-pumping only has a damping of $\alpha=2.8 \times 10^{-3}$. At a frequency of 6\,GHz and an external field $\mu_{0}H$ = 160\,mT they obtain a decay length of $l_{\text{decay}}=3.6\,\mu$m. No values for the magnon lifetime or group velocity are presented.

A very detailed investigation is presented by Qin \textit{et al.}~\cite{Qin2018}. They investigate magnon propagation at various magnetic fields, frequencies, and k-vectors. The results clearly show that due to the dispersion relation (Fig. \ref{Dispersion}), that has the highest slope close to the uniform mode, small k-vectors show a larger group velocity than large ones. Similarly for constant k-vector smaller magnetic fields also yield larger group velocities, mainly because the shape of the dispersion relation changes for different magnetic fields. In addition lower magnetic fields which go along with lower frequency for constant k-vector also lead to an increasing lifetime which is inversely proportional to the frequency (eq. \ref{lifetime}). For a 40\,nm thick film with a damping of $\alpha = 3.5 \times 10^{-4}$ they find a magnon lifetime of up to $\tau$ = 500\,ns and group velocities up to $v_{\text{g}}$ = 3000\,m/s. From these values a decay length of $l_{\text{decay}} = 1.5$\,mm can be inferred.

Based on this analysis also the following two results can be understood. Collet \textit{et al.} \cite{Collet2017} use a 20\,nm thick film and perform their investigation at $f$ = 6.6\,GHz and $\mu_{0}H$ = 150\,mT. For the relatively thin film the group velocity observed is approx. $v_{\text{g}}\approx$ 320\,m/s. The observed decay length is only $l_{\text{decay}} = 25\,\mu$m corresponding to a magnon lifetime of $\tau$ = 78\,ns. Especially the small magnon lifetime can be related to the high frequency and magnetic field which are used.

Yu \textit{et al.} \cite{Yu2014} achieve large magnon lifetimes up to 620\,ns working in the frequency range between 0.7 and 1.5\,GHz. Although the YIG film which is used is only 20\,nm thick they achieve  group velocities up to $v_{\text{g}}$ = 1200\,m/s which decrease to 600\,m/s for higher k-vectors. These values are determined at a relatively low magnetic field of $\mu_{0}H$ = 5\,mT. The maximum propagation length resulting from these properties is $l_{\text{decay}} = 580\,\mu$m.

So even for very thin YIG films it is possible to achieve propagation lengths which are sufficiently large for integrated magnonics. However, it also becomes clear from these results that the propagation length is only of limited use to determine the Gilbert damping. Although the magnon lifetime is directly related to the damping the group velocity strongly depends on magnetic field, frequency and k-vector making the extraction of the damping itself rather complicated.

After discussing the lifetime, it makes sense to have a closer look at the linewidth. Because the linewidth is often measured at a fixed frequency by sweeping the magnetic field we discuss the linewidth in units of the magnetic field.
In an ideal system the linewidth in FMR  $\mu_{0}\Delta H$ is proportional to the frequency and the linewidth for zero frequency $\mu_{0}\Delta H_{0}$ is zero:

\begin{figure}[h]%
\includegraphics[width=\linewidth]{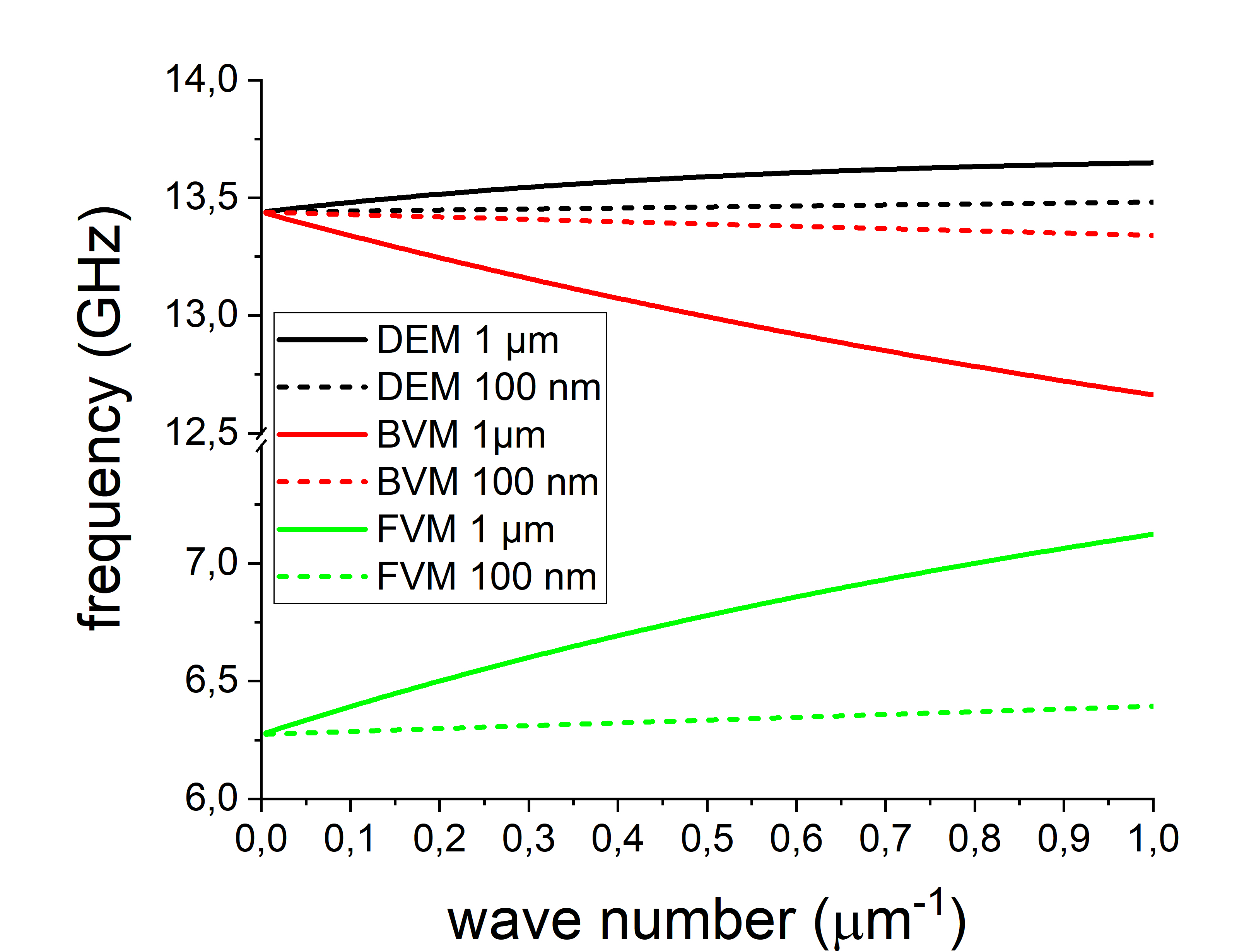}
\caption{Calculated dispersion relation for a 100\,nm thick and a 1\,$\mu$m thick YIG film, respectively, with backward volume mode (BVM), Damon Eshbach mode (DEM), and forward volume mode (FVM). It is clearly visible that $|\partial\omega/\partial k|$ is smaller for the thinner film, at least for small k-vectors.}
\label{Dispersion}
\end{figure}

\begin{equation}\label{linewidth}
\mu_0\Delta H_{\text{FMR$_{\text{HWHM}}$}}=\frac{2\pi\alpha f_{\text{res}}}{\gamma}
\end{equation}

with $\gamma/2\pi = 28$~GHz/T being the gyromagnetic ratio and $\alpha$ being the Gilbert damping. This equation uses the half width at half maximum (HWHM) value of the linewidth. As mentioned before, however, the magnon lifetime is the relevant parameter for magnonics applications.
Using eqs. \ref{linewidth} and \ref{lifetime} we are allowed to directly convert linewidth into lifetime.
The latter is no longer possible if $\mu_{0}\Delta H_{0}\neq0$. Unfortunately a vanishing $\mu_{0}\Delta H_{0}$ is unrealistic and it needs to be understood how $\mu_{0}\Delta H$, $\tau$ and $\alpha$ are interrelated and how this needs to be taken into account when assessing the layer quality.

From eq. \ref{linewidth} we can see that a certain linewidth at a fixed frequency always imposes an upper limit to the damping because it assumes $\mu_0\Delta H_{0}=0$. This upper limit corresponds to a damping of approx. $ 2.8\times10^{-4}$ per $100\,\mu $T linewidth at 10\,GHz.

If $\mu_{0}\Delta H_{0}$ is finite the following equation may be applied:

\begin{equation}\label{alpha1}
\mu_0\Delta H_{\text{FMR$_{\text{HWHM}}$}}=\mu_{0}\Delta H_{0_{\text{HWHM}}}+\frac{2\pi \alpha f_{\text{res}}}{\gamma}
\end{equation}

where this equation uses again the HWHM value of the linewidth.

It should be noted that this equation assumes that the inhomogeneous linewidth broadening $\mu_{0}\Delta H_{0}$ is completely frequency independent, an assumption which is difficult to prove in the experiment (if at all). An inhomogeneous broadening involving two-magnon scattering for example would be frequency dependent. So in this case something intuitively odd can be observed. When for a constant linewidth $\mu_{0}\Delta H$ at a given frequency the inhomogeneous linewidth $\mu_{0}\Delta H_{0}$ is increased the slope of $\mu_{0}\Delta H(f)$ decreases which corresponds to a nominally reduced damping (eq. \ref{alpha1}). This happens despite the fact that we apparently introduce physics which at low frequency broaden the lines and thus seem unfavorable for long lived magnons.

So it is theoretically possible to extract ultra low damping values from very broad lines if the linewidth does (almost) not change with frequency. Nevertheless, the broadening can have multiple reasons. For example the layer can be inhomogeneous or one can observe a line which in reality is composed of several lines and the change in linewidth with changing frequency is merely caused by the frequency dependent line position of the individual lines. For any discussion one also needs to consider that eq. \ref{alpha1} is based on a single spin approximation and it is at least debatable whether this approximation holds for an inhomogeneous sample. It is thus of utmost importance to determine whether the magnon lifetime $\tau_{\text{magnon}}$ is only given by $\alpha$ as determined from eq. \ref{alpha1} or whether the physics leading to a large $\mu_0\Delta H_0$ in FMR can reduce the magnon lifetime, a question which in some cases may only be answered by investigating magnon propagation or time domain FMR.

\begin{figure}[h]%
\includegraphics*[width=\linewidth]{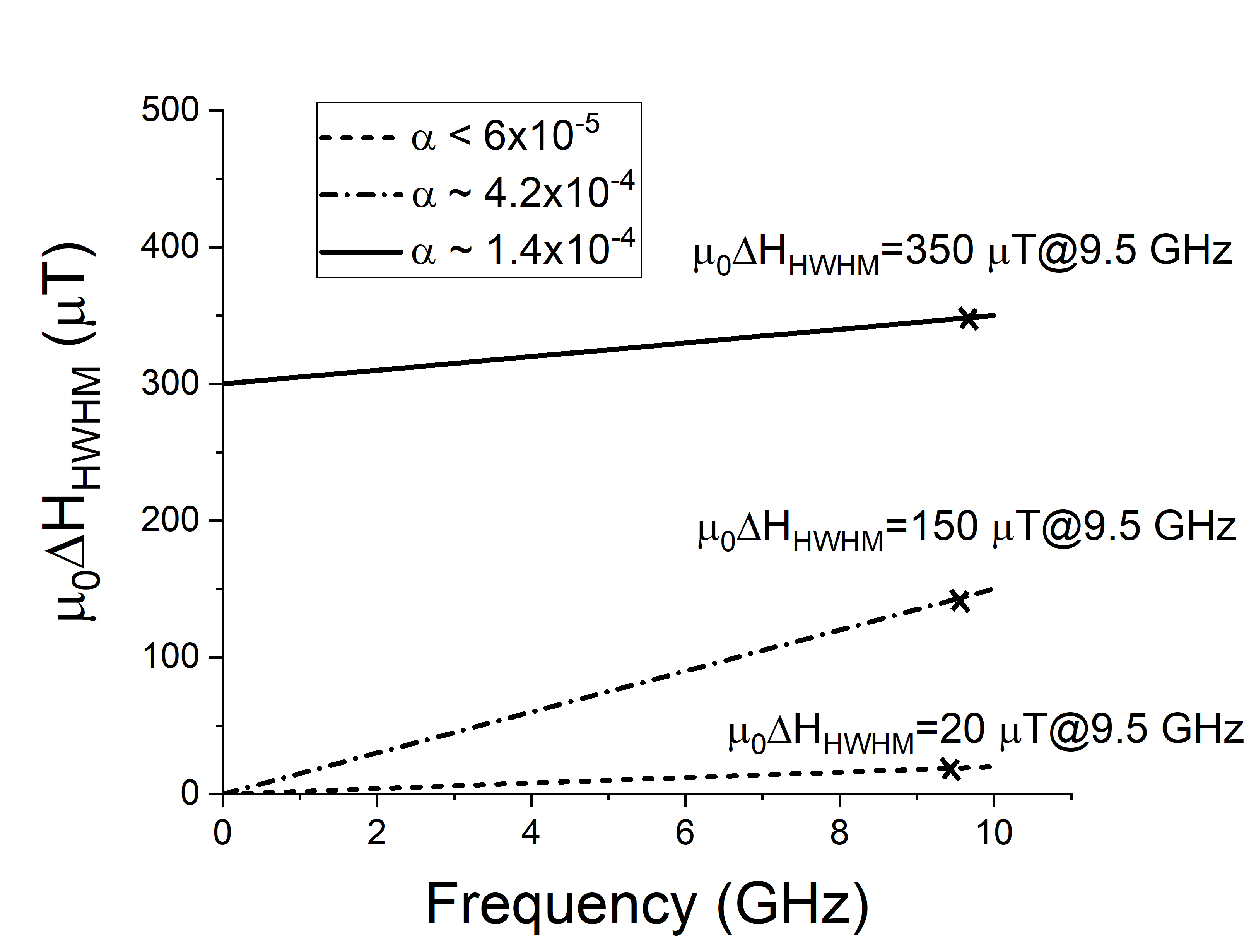}
\caption{Examples for different combinations of linewidth (at 9.5\,GHz) and damping. The dashed line shows a very small linewidth. The corresponding damping is also very small, although it is the maximum possible value for this linewidth because $\mu_0\Delta H=0$. The dash-dotted line shows a much larger damping, although with 150\,$\mu$T the linewidth is still small. For the solid line the damping is again lower, although with 350\,$\mu$T the linewith is much larger. In this case the low damping is only due to the intercept of the plot with the y-axis corresponding to $\mu_0\Delta H_0=300\,\mu$T.  }
\label{Linedamp}
\end{figure}

As a consequence it is realistic to state that a very narrow linewidth is a reasonable indicator for long magnon life times and always must go along with a reasonably low damping. What damping is extracted in the analysis, is then determined by $\mu_0\Delta H_0$ which for a linear dependence of linewidth on frequency also determines the slope. On the other hand, a small damping can be considered a necessary but not always sufficient condition for long lived magnons. Ideally always both values, linewidth and damping should be mentioned.
Furthermore, when the damping is determined by measuring the frequency dependence of the FMR linewidth it is good to check whether properties of a single line and of a homogeneous layer are observed. This can be achieved by for example doing measurements over a wide frequency range and by doing experiments on samples of different size.
For these reasons also our following discussion of thin film YIG the properties will always discuss damping and FMR linewidth (if given in the respective publications).
It should be noted that although the linewidth is mentioned in most of the relevant publications a direct comparison needs to be done with caution. In different experiments the linewidth may be determined in a different way and even the definition of the linewidth can be different. In most cases the relevant linewidth is the half width at half maximum (HWHM). In some cases, also the full width at half maximum (FWHM) may be mentioned which intuitively is more like the width of the line. A third alternative results from the way the ferromagnetic resonance is measured. Quite often the sensitivity of the FMR detection is enhanced by modulating the external magnetic field and using lock-in detection. As a consequence the measured signal is the derivative of the RF-absorption. From this data usually the peak-to-peak (p-p) linewidth is extracted and sometimes listed as the relevant linewidth. There is, however, a conversion factor of $\sqrt{3}/2$ which needs to be applied to the peak to peak value to get to the HWHM. So the p-p value overestimates the linewidth compared to the HWHM. Quite often it is not explicitly if it is not mentioned what definition of the linewidth is used. However, in many cases this can be concluded from the way the linewidth is converted into a damping. So while eq. \ref{alpha1} is to be used with HWHM the relevant equation for the p-p linewidth would be

\begin{equation}\label{alpha2}
\mu_0\Delta H_{\text{FMR$_{\text{p-p}}$}}=\mu_0\Delta H_{0_{\text{p-p}}}+\frac{2}{\sqrt{3}}\frac{2\pi \alpha f_{\text{res}}}{\gamma}
\end{equation}

and for FWHM

\begin{equation}\label{alpha3}
\mu_0\Delta H_{\text{FMR$_{\text{FWHM}}$}}=\mu_0\Delta H_{0_{\text{FWHM}}}+\frac{4\pi \alpha f_{\text{res}}}{\gamma}
\end{equation}

From the changing prefactor one can also deduce the meaning of $\mu_0\Delta H_{0}$ in the equation which in eq. \ref{alpha1} would be $\mu_0\Delta H_{0_{\text{HWHM}}}$ while in eq. \ref{alpha2} it means $\mu_0\Delta H_{0_{\text{p-p}}}$ and in eq. \ref{alpha3} $\mu_{0}\Delta H_{0_{\text{FWHM}}}$.

Finally it should me mentioned that for the extraction of the linewidth a fit by a Lorentzian line shape should be done in order to make sure that indeed a single line is present and not multiple overlapping lines.

\section{Growth methods}
In the following, three different growth methods are described and compared in terms of results, namely high temperature PLD, room temperature sputtering with subsequent annealing, and room temperature PLD with subsequent annealing. For all three of them, growth of sub-100\,nm YIG films with extraordinary properties has been published some with outstanding results and several more which still can be considered as high quality films. It is noteworthy that in all of these experiments gallium gadolinium garnet (GGG) has been used as a substrate. Besides the fact that the garnet structure of GGG is favorable for YIG growth, the lattice constant of the GGG substrate (1.2383\,nm) is very close to the YIG bulk value of 1.2376\,nm. As expected, most layer grow pseudomorphic and are fully strained.
Although most experiments use (111) orientation a few exceptions show that the crystalline orientation of the substrate does not have a clear influence on the magnetic layer quality, see Table 1. In order to place the values given below into the right context it should be noted that for thick films grown by LPE a FWHM linewidth of 15\,$\mu$T\,@\,9.5\,GHz was reported \cite{Glass1976} which as described above corresponds to an upper limit for the damping of $\alpha=2\times10^{-5}$.

For sake of completeness in Table 1 we also list the values of the determined saturation magnetization in the various experiments which can be compared to the bulk value for YIG which is $\mu_{0}M_{\text{S}}$  $\approx$\,180\,mT~\cite{Hansen1974}. However, it turns out that although there is a huge variation of $\mu_0M_{\text {S}}$ no clear relation between the saturation magnetization and the damping can be observed. Also the values need to be taken with a grain of salt. In some cases the saturation magnetization is obtained by a fit to the Kittel formula~\cite{Kittel1948} which leaves an uncertainty of the crystalline anisotropy which in some cases may not be negligible \cite{Hauser2016}. In those cases $\mu_{0}M_{\text{eff}}$ will be given instead.

\subsection{Pulsed Laser Deposition}
High temperature PLD was used for the fabrication of thin film YIG already in the nineties \cite{Dorsey1993}. Already in 1993 the growth of YIG on GGG by PLD was demonstrated. In that case no value for the damping was determined. Nevertheless, a linewidth of $\mu_0\Delta H_{0_{\text{FWHM}}} = 100\,\mu$T$\,@\,9.5$\,GHz which sets the damping to $\alpha \leq 1.5 \times 10^{-4}$ assuming $\mu_0\Delta H_{0}=0$. Obviously the real damping is even lower than this, most likely below $1 \times 10^{-4}$. In this case, the thickness of the layer was 1\,$\mu$m which is above the range that we are discussing here. Nevertheless, as an early and extraordinary result it can serve as a reference of what is possible using PLD. Similar results were obtained later by Manuilov \cite{Manuilov2009} who achieved $\mu_{0}\Delta H = 90\,\mu$T$\,@\,9.5$\,GHz for a 1.22\,$\mu$m thick film, however, only for measurements with the magnetization saturated perpendicular to the film. This measurement geometry can avoid two magnon scattering and in many cases can yield damping values below those measured with in-plane magnetization \cite{Mcmichael2004}. As mentioned both these experiments used film thicknesses much larger than 100\,nm and more results can be found in literature for this thickness range.

\begin{figure}[h]%
\includegraphics*[width=\linewidth]{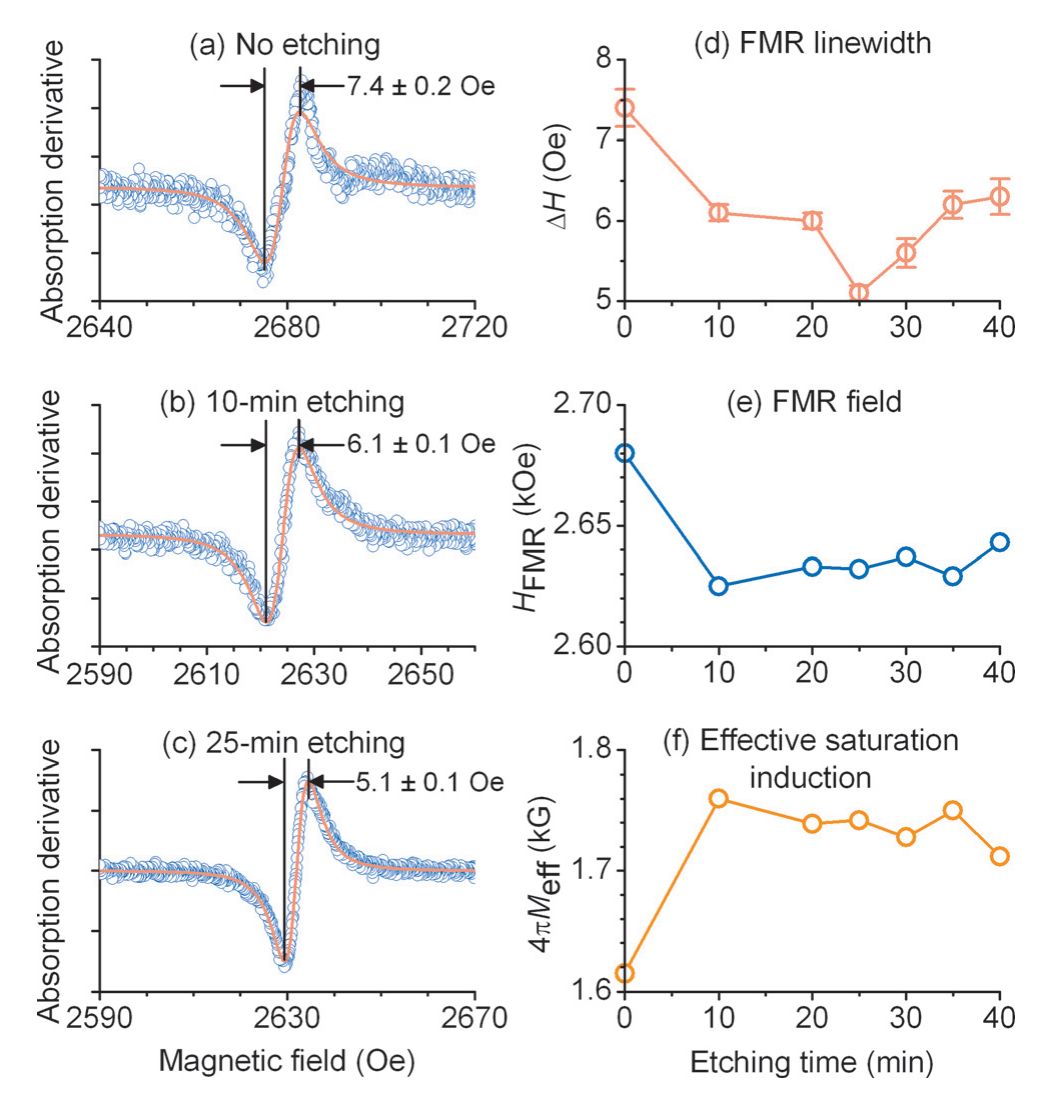}
\caption{Reproduced from Sun \textit{et al.}, Applied Physics Letters 101(15), 152405 (2012)~\cite{Sun2012}, with the permission of AIP Publishing. (a)-(c) FMR profiles for an 11\,nm film before and after etching, as
indicated. (d)-(f) FMR linewidth, FMR field, and effective saturation induction as a function of etching time.}
\label{Sun4}
\end{figure}

For thinner films grown at high temperature a first outstanding result was published in 2012 by Sun \textit{et al.}~\cite{Sun2012}. In this publication a p-p linewidth of $\mu_0\Delta H_{\text {p-p}} = 340\,\mu$T $@9.5$\,GHz was achieved for an as-grown YIG film of 19\,nm thickness grown by high temperature PLD on (111)-oriented GGG substrates. Because $\mu_0$$\Delta H_{0}$ is almost 300\,$\mu$T the damping is also very small with $\alpha=2.3 \times 10^{-4}$. In this work the correlation of growth conditions, structural and chemical properties of the layers and damping is investigated. After deposition an annealing step in oxygen at growth temperature for 10\,min is performed. The authors observe that low growth rate and higher substrate temperature (maximum substrate temperature is $T_{\text{S}}=850^{\circ}$C) yield the smallest linewidth. They attribute this mainly to the lowering of the surface roughness and Fe deficiency at the surface which decreases for increasing growth temperature and decreasing growth rate. The Fe deficiency and its change with growth parameters is evidenced by XPS. The Fe deficiency and the roughness should increase two magnon scattering and thus cause the linewidth broadening. This theory is further supported by removal of the top surface by soft ion bombardment which indeed within certain limits reduces the FMR linewidth of a sample, Fig.~\ref{Sun4}. The surface roughness of the samples as-grown was as small as 0.16\,nm. Based on these results the authors also claim that the trend of higher damping for thinner films which is also observed by other groups can be attributed to the decreasing contribution of the bulk compared to the detrimental influence of the surface layer. No lattice constant of the YIG layers is explicitely mentioned but the X-ray diffraction shows a larger interlayer distance for the strained YIG film than for the GGG substrate meaning that in the relaxed state the YIG would also have a larger lattice constant than the GGG. At first glance this is surprising because the lattice constant of bulk YIG is smaller than the one of GGG. However, this effect is observed for all experiments discussed below (as far as can be determined from the information supplied).
The two layers for which $\mu_{0}M_{\text{eff}}$ is listed show values of 167\,mT (11\,nm thickness) and 188\,mT (19\,nm thickness). The second value is even above the bulk value. As shown in the following, a trend of larger $M$ for thicker films is often observed.

\begin{figure}[t]%
\includegraphics*[width=\linewidth]{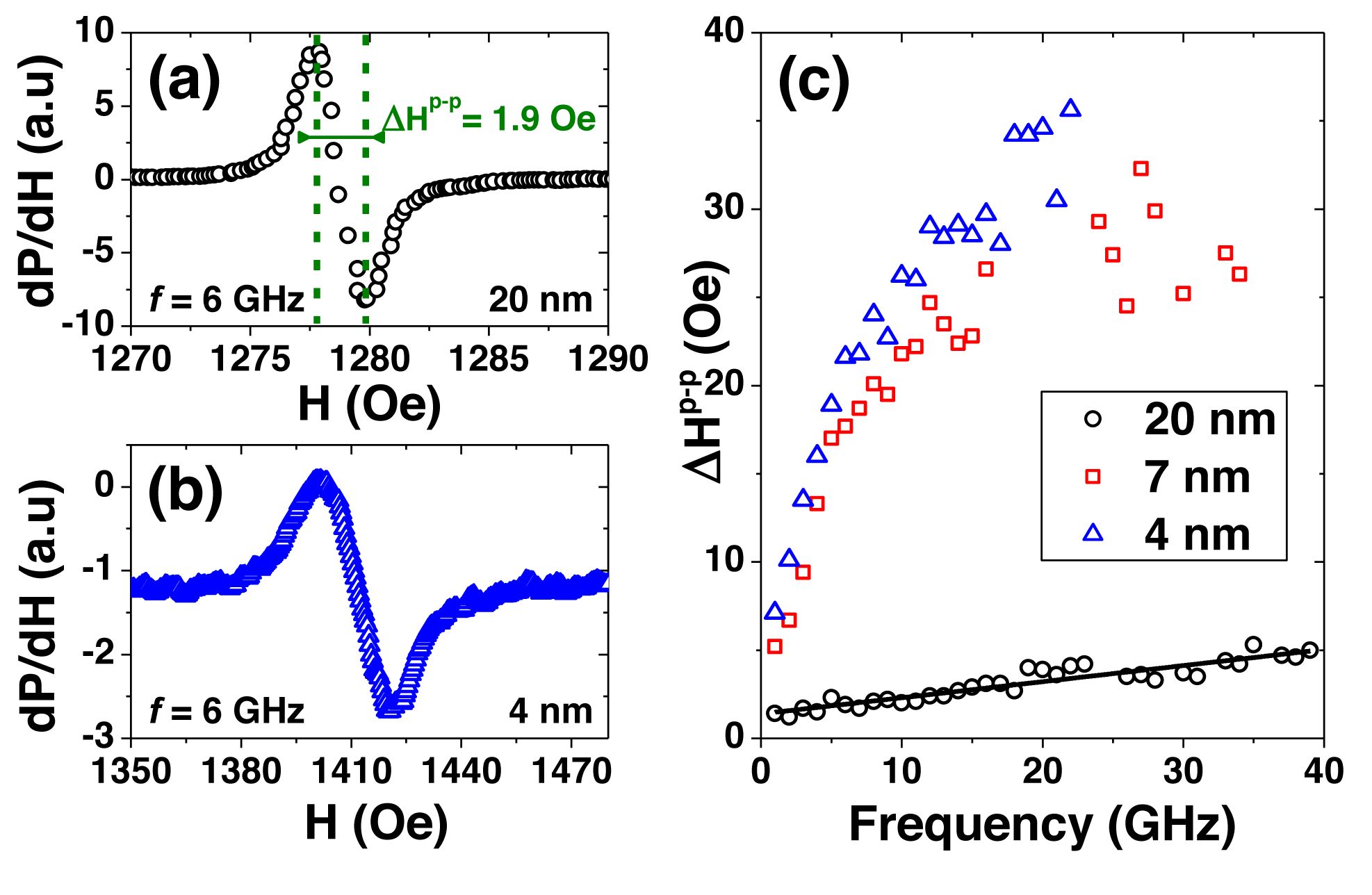}
\caption{Reproduced from d'Allivy Kelly \textit{et al.}, Applied Physics Letters 103(8), 082408 (2013)~\cite{Kelly2013}, with the permission of AIP Publishing. (a) and (b) FMR absorption derivative spectra of 20 and 4\,nm thick YIG films at an excitation frequency of 6\,GHz. (c) rf excitation frequency dependence of FMR absorption linewidth measured on different YIG film
thicknesses with an in-plane oriented static field. The black continuous line is a linear fit on the 20\,nm thick film from which a Gilbert damping coefficient of $2.3 \times 10^{-4}$ can be inferred ($\Delta H_{\text p-p} = \Delta H_{0}+\alpha\frac{4\pi}{\gamma\sqrt{3}}f$). The damping of the 7\,nm and 4\,nm films is significantly larger but most off all the frequency dependence is not linear.}
\label{Kelly2}
\end{figure}

In 2013 d'Allivy Kelly and coworkers published an even smaller linewidth for similarly thin YIG films \cite{Kelly2013}. Here, layers with three different respective thicknesses were investigated (4\,nm, 7\,nm, 20\,nm). Even the thickest of those layers is among the thinnest for which a linewidth of approx. 200\,$\mu$T has ever been reported. The p-p linewidth at 6\,GHz for the 20 nm film is as small as 190\,$\mu$T, Fig. \ref{Kelly2}. For comparison: taking into account the extracted damping this corresponds to a HWHM of approx. 200\,$\mu$T at 10 GHz. The corresponding value for the damping is $\alpha=2.3\times10^{-4}$ which is larger than for thicker films but up to that time together with the result from Sun \textit{et al.} the best value observed for that thickness range. The fact that despite the smaller linewidth compared to Sun \textit{et al.} the damping is not decreased is due to the smaller zero frequency linewidth. The layers were deposited on (111)-oriented GGG substrates at a temperature of 650\,$^\circ$C. It is interesting to note that the substrate temperature is much lower than the optimum value determined by Sun \textit{et al.} and the linewidth is achieved despite a surface roughness of 0.23\,nm which is larger than for the aforementioned layers. The cubic lattice constant of the 20\,nm thick film is 1.2459\,nm and thus considerably larger than that of the substrate. While for a thickness of 20\,nm the layer quality is excellent, for the thinner films presented by d'Allivy Kelly \textit{et al.}, however, the magnetic properties quickly start to degrade. For these films $\mu_{0}M_{\text{S}}$ was determined by SQUID magnetometry. With a value of 210\,mT for films of 20\,nm and 7\,nm thickness the values are much higher than the bulk value and only for 4\,nm thickness $\mu_0M_{\text S}$ goes down to 170\,mT. These results are exceptional because in all experiments discussed here $\mu_{0}M_{\text S}$ is not or only slightly larger than in bulk YIG. In a subsequent publication \cite{Yu2014}, a magnon decay length of ~580\,$\mu$m was estimated for a film with identical properties fabricated by the same process.

\begin{figure}[h]%
\includegraphics*[width=\linewidth]{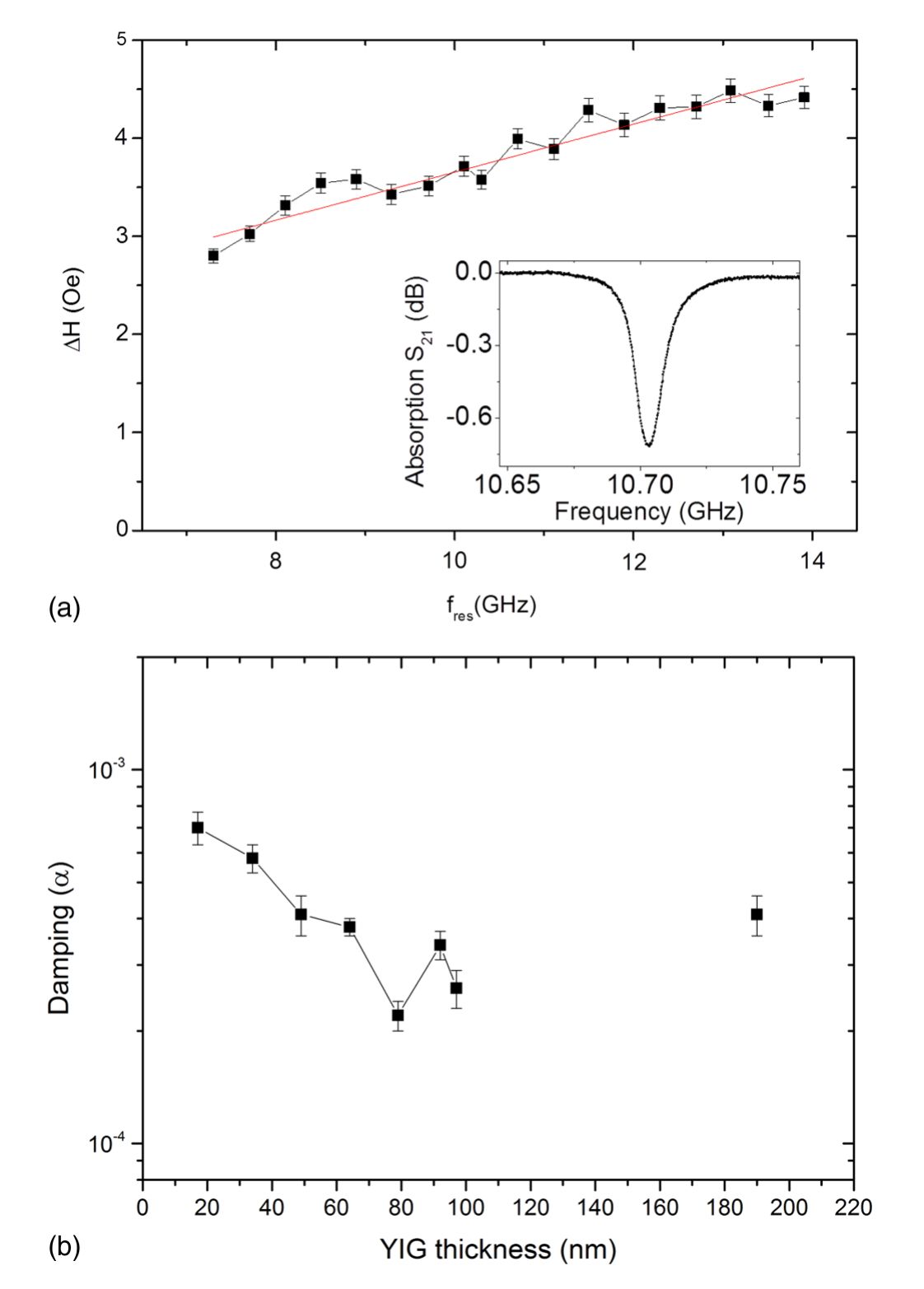}
\caption{Reproduced from  Onbasli \textit{et al.}, APL Materials 2(10), 106102 (2014) \cite{Onbasli2014}, with the permission of AIP Publishing. (a) $\Delta H$ as a function of the resonance frequency. The inset shows one example of the measurement for a resonance frequency of 10.704\,GHz. All these data were recorded for the 92\,nm thick YIG film.
Damping ($\alpha$) is $3.4\times10^{-4}$ and $\Delta H_{0}$ = 1.2\,Oe for this sample. (b) Damping parameter of YIG films as a function of film thickness.}
\label{Onbasli4}
\end{figure}

One year later low damping and linewidth for PLD grown thin YIG films were also reported by Onbasli \textit{et al.}~\cite{Onbasli2014}. Among the results discussed here it is the only one where (001)-oriented GGG substrates were used. Here a linewidth of 300\,$\mu$T at 10\,GHz is mentioned for a 79\,nm thick layer, however, the FWHM is given. So with $\mu_0\Delta H_{\text{HWHM}}= 150\,\mu$T the linewidth is even lower than for d'Allivy Kelly~\cite{Kelly2013}. The corresponding damping value is $\alpha=2.2 \times 10^{-4}$. Onbasli \textit{et al.} investigate several thicknesses and also observe a clear trend towards higher damping for thinner films (Fig.~\ref{Onbasli4}). For comparison a 34\,nm film prepared by the same method exhibits $\alpha = 5.8\times10^{-4}$. As in \cite{Kelly2013} the films were deposited at a substrate temperature of 650\,$^\circ$C. For layer thicker than 50 nm the out-of-plane lattice constant was determined by X-ray diffraction, which varied from 1.2391\,nm to 1.2408\,nm without a clear dependence on the thickness. For layers with thicknesses between 34\,nm and 190\,nm thickness $\mu_{0}M_{\text S}$ varied between 167\,mT and 172\,mT with no systematic thickness dependence. Only for a very thin film (17\,nm) $\mu_{0}M_{\text S}$ dropped to 158\,mT.

\begin{figure}[h]%
\includegraphics*[width=\linewidth]{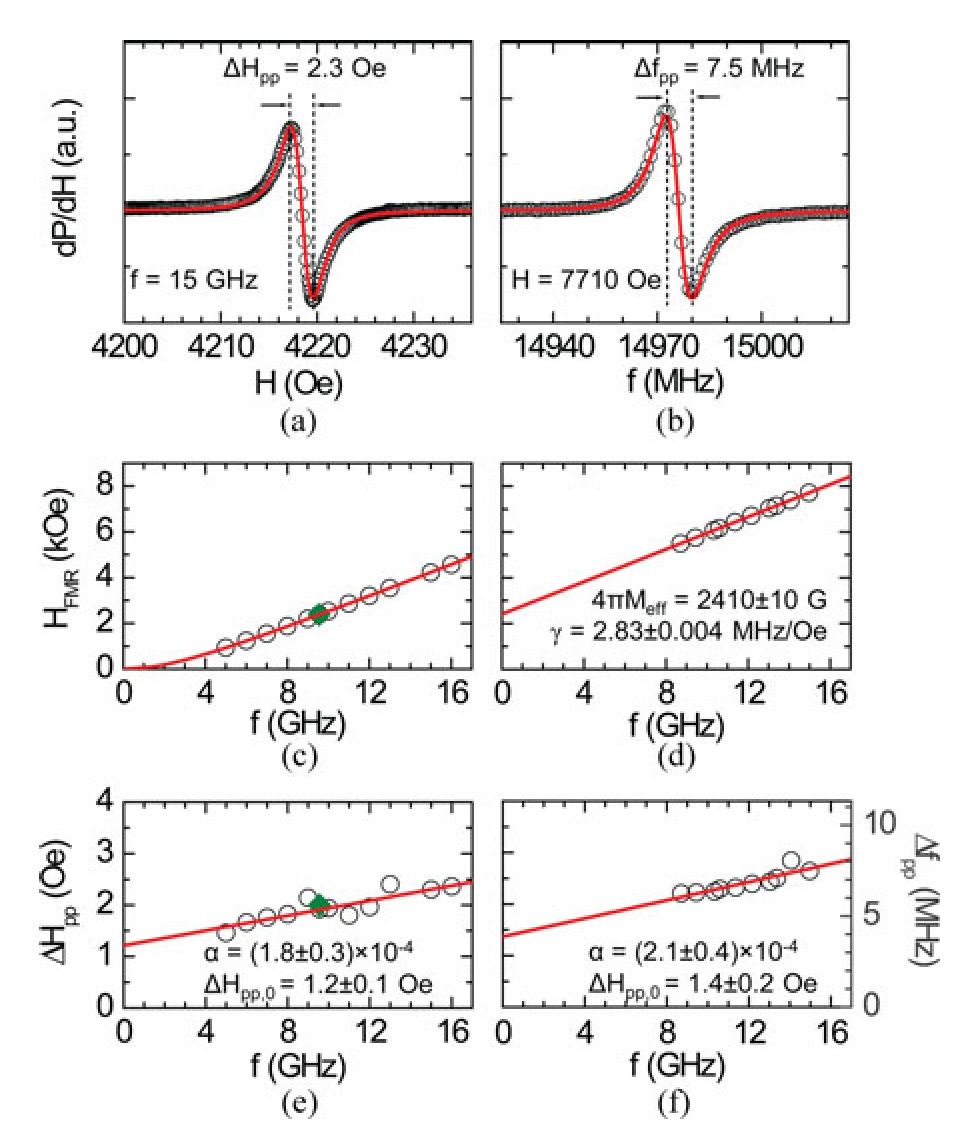}
\caption{\textcircled{c} [2015] IEEE. Reprinted, with permission, from Howe \textit{et al.} IEEE Magnetics Letters 6, 3500504 (2015)~\cite{Howe2015}. (a, b) FMR spectra measured with a broadband FMR system: (a) field-sweep spectrum with in-plane bias field at fixed $f$ = 15\,GHz, and (b) frequency-sweep spectrum with fixed out-of-plane bias field $H$ = 7710\,Oe. (c,d) Resonance field $H_{\text{FMR}}$ versus microwave frequency for in-plane bias field (c) and out-of-plane bias field (d). (e, f) Peak-to peak linewidth $\Delta H_{\text{p-p}}$ versus microwave frequency f for in-plane bias field (e) and out-of-plane bias field (f). All data are measured from a 23\,nm thick YIG film. Solid green symbols in (c) and (e) indicate values measured in a cavity operated at $f$ = 9.56\,GHz.}
\label{Howe4}
\end{figure}

In 2015 Howe \textit{et al.} also published data from 23\,nm thick PLD deposited films~\cite{Howe2015}. In this case the damping was even lower than for~\cite{Kelly2013} with $\alpha\approx1.8\times10^{-4}$ with slight variations depending on the measurement technique. With $\mu_0\Delta H_{\text{p-p}}=200\,\mu$T the linewidth at 10 GHz is also smaller than that reported in \cite{Kelly2013}, Fig. \ref{Howe4}. Howe and coworkers used a higher deposition temperature of 825\,$^\circ$C. The lattice constant that they obtained by x-ray diffraction (1.2525\,nm) is even larger than that observed in \cite{Kelly2013}. $\mu_{0}M_{\text S}$ was measured by vibrating sample magnetometry (VSM) to 160\,mT.

Another interesting result with low damping was obtained by Tang \textit{et al.} \cite{Tang2016} in 2016. They deposited YIG films on (110)-oriented substrates at a temperature of 750$^\circ$C. By annealing the substrates at high temperature prior to growth they achieve a very smooth substrate surface and as a consequence also very flat layers. With 0.067\,nm the root mean square (RMS) value of the surface roughness is the smallest for high temperature PLD grown films discussed here. Damping and linewidth of the films show a peculiar dependence. For a thin film of 17\,nm thickness the authors observe a damping of $\alpha = 7.2 \times 10^{-4}$, while for a 100\,nm thick film the damping is $\alpha = 1.0 \times 10^{-4}$ Nevertheless, for the thick film the zero frequency linewidth is $\mu_0$$\Delta H_{0}\approx 700\,\mu$T while for the thin film it is $\mu_0$$\Delta H{_0}\approx 250\,\mu$T. Even up to a frequency of 8\,GHz the linewidth for the thin film remains below the one measured for the thick film although the damping of the thick film nominally is 7 times smaller. Still the observation of Sun \textit{et al.} \cite{Sun2012} seems to be confirmed that low surface roughness is a necessary ingredient for low damping. One more result from this work should be mentioned. For the films grown on (110)-oriented substrate hysteresis loops taken by vibrating sample magnetometry at room temperature indicate a large in-plane anisotropy and in certain in-plane directions almost 20\,mT are necessary to fully saturate the films while for films grown on (111)-surfaces the anisotropy is usually only a few hundred $\mu$T. Tang \textit{et al.} do not mention an explicit value for $\mu_{0}M_{\text S}$ but from a hysteresis loop a value of approx. 180\,mT can be extracted which is close to the bulk value.

Further results were reported in the last five years, but for PLD at elevated temperatures to the best of our knowledge none of them show similar combinations of low damping and small linewidth. Nevertheless, we will discuss the best results known to us.

In 2014 Hahn \textit{et al.} \cite{Hahn2014} reported on 20 nm thick YIG films and nanostructures fabricated by the same process as used in \cite{Kelly2013}. The nanopatterning was performed by electron beam lithography and ion-milling. Interestingly the damping that is obtained for the continuous film is only $\alpha = 4\times10^{-4}$ compared to $\alpha  =2.2\times10^{-4}$ in \cite{Kelly2013} for a similar film nominally fabricated by the same process. For a nanodisk of 700\,nm diameter, no damping is given, however, the linewidth at 8.2\,GHz is smaller than for the continuous film. It should, however, be noted that the FMR measurements on the disk were done in a perpendicular bias field while those for the film were performed with in-plane field. Depending on the layer the measurement in perpendicular field can yield a smaller linewidth than for the in-plane field due to the absence of two-magnon scattering as mentioned in \cite{Sun2012} and also observed by Manuilov \textit{et al.} \cite{Manuilov2009}. Nevertheless, the linewidth observed for the disk is so small that in the limit of $\Delta H_{0}=0$ the damping of the disk cannot be larger than the one determined for the film. $\mu_{0}M_{\text S}$ of the film is 210\,mT as in~\cite{Kelly2013}.

In 2017 two more papers appeared by Collet and coworkers in which also 20\,nm thick films fabricated as described in \cite{Kelly2013} are used. In one of them, the layers show a damping of $\alpha=4\times10^{-4}$ with $\mu_{0}M_{\text{eff}}$ of 213\,mT \cite{Collet2017}, in the other one the damping is $\alpha = 4.8\times10^{-4}$ \cite{Collet2017_AIP}.

In 2015 another publication by Jungfleisch \textit{et al.} \cite{Jungfleisch2015} reported damping values for layers grown by the process presented in 2014 by  Onbasli \textit{et al.} \cite{Onbasli2014}. Also in this case the damping obtained is not as good as the one reported in the original paper. The best value is achieved for a 75\,nm thick film with a value of $\alpha = 4.89\times10^{-4}$ compared to $\alpha = 2.2\times10^{-4}$ published in \cite{Onbasli2014}. Here the values for $\mu_0M_{\text S}$ partly differ from those obtained in \cite{Onbasli2014}. For the 75\,nm film they observe $\mu_{0}M_{\text S}$ = 166\,mT which is close to the value of the original paper. For a 20\,nm film, however $\mu_{0}M_{\text{S}}$ drops dramatically to 103\,mT.

A final example from 2018 used YIG thin films for the investigation of spin wave propagation. Qin \textit{et al.}\cite{Qin2018} also deposit the films by PLD on (111)-oriented GGG substrates. The deposition temperature is 800\,$^\circ$C and the layer thickness is 40\,nm.  With 144\,mT $\mu_{0}M_{\text S}$ is considerably smaller than the bulk value. X-ray diffraction again shows the YIG lattice constant to be larger than that of the substrate.

Summarising these results, it is obvious that depending on the process details it is possible to obtain high quality YIG at least in a substrate temperature range from 650\,$^\circ$C to 850\,$^\circ$C. Although most publications present growth on (111)-oriented GGG substrates, low damping was also obtained for films grown on (011) or (001)-oriented GGG. For (011), however, the linewidth was relatively large. For all films the lattice constant is larger than for bulk YIG and even larger than for the GGG substrate. In view of these results the assumption by Sun \textit{et al.} that surface roughness and damping are related are at least not disproved. All films with low damping also have a very low surface roughness.
The second assumption of a change in stoichiometry at the surface, however, is debatable. None other of the publications presents XPS data. Some at least show X-ray rocking curves with very small FWHM indicating a homogeneous lattice constant throughout the film. As we will show later, however, Fe deficiency at the surface can go along with a narrow peak in the rocking curve. However, this Fe deficiency is even observed for the films with the lowest damping demonstrated so far.

Besides this, neither lattice constant nor saturation magnetization can be correlated to the magnetization dynamics. Among layers with high quality the values of $\mu_{0}M_{\text S}$ vary between 160\,mT and 210\,mT, the first value well below, the second well above the bulk value. The only clear trend is that for the thinnest films the magnetization seems to drop rapidly, however in some cases this drop happens already at 20\,nm thickness, in other cases well below this value.

\subsection{Off-axis sputtering} A special technique for growth of thin film YIG mainly developed by F. Yang and presented for example in~\cite{Yang2018} is off-axis sputtering. The films which are grown by this technique distinguish themselves from other results in several ways. The saturation magnetization for a 160\,nm thick film was determined to $\mu_{0}M_{\text S}$ = 202\,mT which is even higher than the bulk value. Also the X-ray diffraction data is different from typical results for sputtering or PLD growth. In \cite{Wang2013} the group also presents X-ray diffraction data for various film thicknesses. They find that depending on film thickness the respective lattice constant perpendicular to plane can be either larger than that of GGG or smaller. This is not consistent with a layer of constant composition and in different states of relaxation but actually can only be explained by different bulk lattice constants. FMR linewidth is given for a 30 nm thick film as $\mu_0\Delta H_{\text{p-p}}=275\,\mu$T corresponding to $\mu_0\Delta H_{\text{HWHM}}=238\,\mu$T. Unfortunately no Gilbert damping is extracted for this film thickness. Nevertheless, based on our initial discussion we can state that even for an unrealistically small inhomogeneous linewidth of $\mu_0\Delta H_0=0$ the damping must be smaller than $\alpha=6.7\times10^{-4}$ and in reality a damping in the lower 10$^{-4}$ range can be assumed. For a 16\,nm thick film the damping is determined to $\alpha=6.1\times10^{-4}$. In this case the linewidth at 10\,GHz is closer to $\mu_0\Delta H_{\text{p-p}}=600\,\mu$T and the inhomogeneous linewidth is $\mu_0\Delta H_{0_{\text{p-p}}}\approx360\,\mu$T. Besides these values the layers show exceptional efficiency in spin pumping experiments. The ISH-voltages measured using even 20\,nm thin YIG films covered by metals with large spin orbit coupling can be as high as 5\,mV. This allows to determine the ISHE not only for materials like Pt or W but also for materials with low SOC as for Cu, Ag, or Ti~\cite{Yang2018}.

\subsection{Room temperature sputtering} 
\begin{figure}[h]%
\includegraphics*[width=\linewidth]{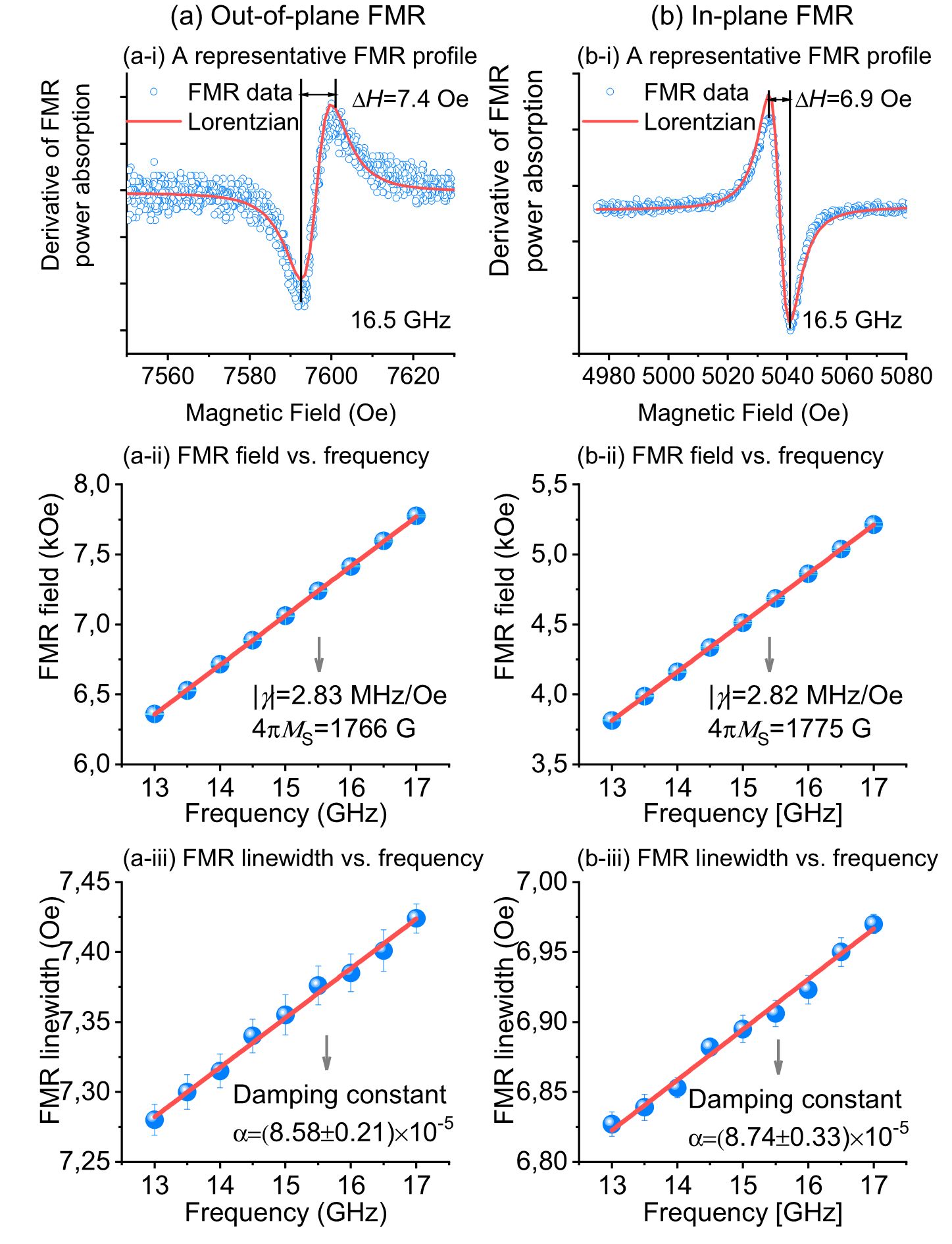}
\caption{\textcircled{c} [2014] IEEE. Reprinted, with permission, from Chang \textit{et al.,} IEEE Magnetics Letters 5, 6700104 (2014)~\cite{Chang2014}. FMR data obtained with the same YIG film cited in Fig. 1 (of Chang \textit{et al.}~\cite{Chang2014}, layer thickness is 22.3\,nm). The left and right columns show the FMR data measured with a static magnetic field applied perpendicular to and in the plane of the YIG film, respectively.}
\label{Chang2}
\end{figure}

To the best of our knowledge the first publication which mentions high quality thin film YIG by room temperature sputtering and subsequent annealing is by Chang \textit{et al.} in 2014 \cite{Chang2014}. At the same time it also reports the lowest damping not only for layers fabricated this way but also for any sub-100\,nm YIG film. The layers were obtained by sputtering YIG from a stoichiometric target using ordinary magnetron sputtering in Ar atmosphere. Subsequently the films were annealed in pure oxygen at a temperature of 800\,$^\circ$C for 4 hours. All films were grown on (111)-oriented GGG substrates. The resulting films are distinguished by a very low surface roughness. The lattice constant is again larger than for GGG. Five different layers grown at different Ar-pressures are presented, all of them with a damping of $\alpha\le3.1\times10^{-4}$. The best value is obtained for a 22\,nm thick film with a damping of $\alpha\le8.6\times10^{-5}$. This record value goes along with a surface roughness of 0.13\,nm. FMR results are presented for perpendicular and in-plane orientation of the external field, respectively. For both cases similar values for the damping are obtained ($\alpha\le8.58\times10^{-5}$ for perpendicular field and $\alpha\le8.74\times10^{-5}$ for in-plane field), Fig.~\ref{Chang2}. Surprisingly the relatively large p-p linewidth of approx. 670\,$\mu T$ at 10 GHz (extrapolated from Fig.~\ref{Chang2}) is even bigger for perpendicular field indicating little influence of two magnon scattering which may be related to the low surface roughness. The authors attribute the high quality to the optimization of the Ar pressure during deposition. For the best film they obtain $\mu_0M_{\text{S}}$ = 177\,mT which is identical to the bulk value of YIG within the measurement uncertainty.

In 2017 more experiments on films grown by the same parameters were reported \cite{Chang2017}. All layers described there were also of a thickness of approx. 20\,nm. Four out of six films showed a damping of $\alpha <2\times10^{-4}$, three out of those four even had a damping below $10^{-4}$ reproducing the former results. The FMR linewidth of these films, however, was even bigger than reported in \cite{Chang2014}. While in 2014 Chang \textit{et al.} reported  690 $\mu$T at 16.5\,GHz the linewidth at 16.5\,GHz for the films reported in 2017 ranges from 850 to 1200 $\mu$T. The values of $\mu_{0}M_{\text {eff}}$ for all films vary between 175\,mT and 192\,mT so on average they are slightly above the bulk value.

Already in 2014 another group also published data on films grown by room temperature sputtering and post annealing~\cite{Lustikova2014}. The films were grown by on-axis sputtering on (111)-oriented GGG substrates and annealed in Air for 24 hours. The surface roughness is extremely small with an RMS value of 0.008\,nm. The interface between YIG and substrate, however, had a much larger roughness of 0.6\,nm and also transmission electron microscopy showed a large number of spherical defects with a diameter of 10\,nm or more, not observed by other groups. The smallest linewidth obtained at 9.5\,GHz is as small as $\mu_0\Delta H_{\text p-p} = 380\,\mu$T but for a broader range of samples a linewidth between 400\,$\mu$T and 600\,$\mu$T is mentioned. The best damping value obtained in this case is $\alpha\le7.0\times10^{-4}$.

In 2017 Talalaevskij et al \cite{Talalaevskij2017} also reported growth of high quality thin YIG films by room temperature sputtering and annealing in air. The annealing was done at T = 850$^\circ$C for 120 minutes. Unfortunately no X-ray diffraction data or lattice constant are supplied, however, X-ray reflectometry shows extremely smooth interfaces and surfaces. The films which are investigated vary in thickness between 19 and 49\,nm. In contrast to other groups magnetization measurements indicate a 4-6\,nm thick magnetically dead layer at the interface to the substrate leading to very low $\mu_0M_{\text S}$ = 100\,mT for the 19\,nm thick film. $\mu_0M_{\text S}$ increases with thickness over $\mu_0M_{\text S}$ = 140\,mT (29\,nm), $\mu_0M_{\text S}$ = 150\,mT (38\,nm), to $\mu_0M_{\text S}$ = 160\,mT for a 49\,nm thick film. Assuming a dead layer of 5\,nm the last value corresponds to the magnetic part of the layer reaching the magnetization value of  bulk material. With $\alpha=2.4\times10^{-4}$ for the 49\,nm thick layer the damping is not as good as reported by Chang \textit{et al.} \cite{Chang2014}, however, the linewidth at 10\,GHz is smaller with $\mu_0\Delta H_{\text{HWHM}} \approx 400\,\mu$T. A similar damping ($\alpha = 2.6\times10^{-4}$) is achieved for 38\,nm thickness while for 29\,nm ($\alpha = 5.8\times10^{-4}$) and 19\,nm film thickness ($\alpha = 8.1\times10^{-4}$) the damping is considerably increased following the trend observed by most groups.

Especially the publications by Chang \textit{et al.} show that room temperature sputtering and annealing can reproducibly deliver YIG thin films with extremely low damping, although the linewidth for these low damping films is typically much bigger than for high quality PLD grown samples. There is, however, no-other group that has published similarly good layers fabricated by the same method.

\subsection{Room temperature PLD} 
\begin{figure}[h]%
\includegraphics*[width=\linewidth]{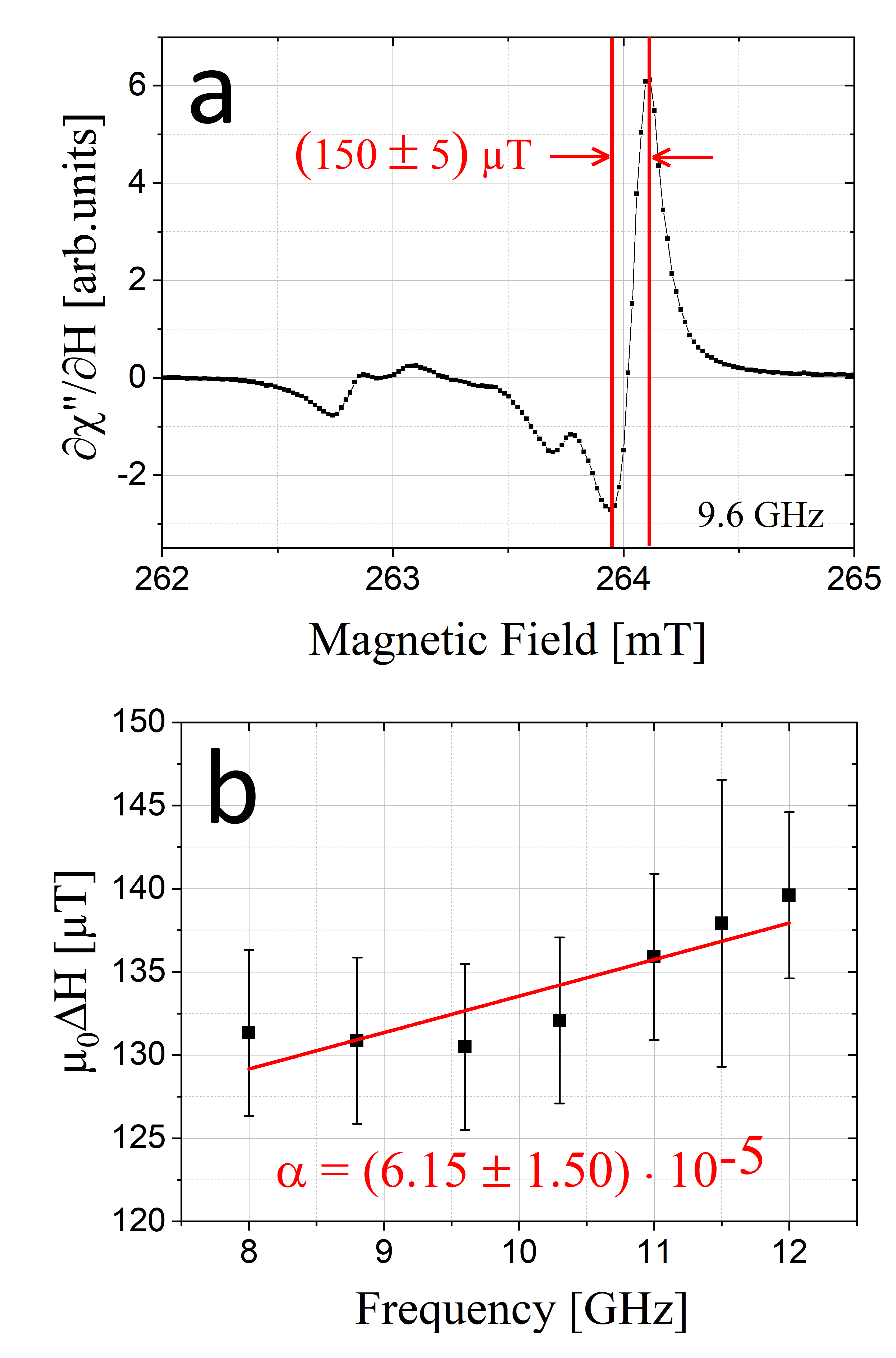}
\caption{Reproduced from Hauser \textit{et al.}, Scientific Reports 6, 20827 (2016)~\cite{Hauser2016} (a) FMR data obtained at 9.6\,GHz for a 56\,nm thick YIG layer after annealing. The main resonance line has a peak-to-peak linewidth of 150\,$\pm$\,5\,$\mu$T. This peak-to-peak linewidth corresponds to a true linewidth of 130\,$\pm$\,5\,$\mu$T. (b) Frequency dependence of the FMR linewidth for the same sample. The fit is a straight line corresponding to a damping of $\alpha$\,=\,(6.15\,$\pm$\,1.50)\,$\times$\,10$^{-5}$.}
\label{Hauser}
\end{figure}

In 2016 results were published from our group on YIG thin films deposited at room temperature by PLD on (111)-oriented GGG substrates~\cite{Hauser2016}. Transmission electron microscopy on the deposited layers showed that they were almost amorphous. The absence of magnetism was confirmed by SQUID magnetometry within the measurement limits. After annealing at temperatures between 800\,$^\circ$C and 900\,$^\circ$C in pure oxygen at ambient pressure the layers became ferrimagnetic with a saturation magnetization which is typically 10\% smaller than the bulk value known for YIG. After annealing, transmission electron microscopy shows an apparently flawless film. No defects are visible, neither in the layer nor at the GGG/YIG interface. The lattice constant is again larger than for the GGG substrate with a lattice mismatch of approx. 0.06\% to the GGG substrate. The surface roughness determined by X-ray reflectometry is at least better than 0.2\,nm. An X-ray rocking curve shows a FWHM for the YIG (444) diffraction peak of 0.015$^\circ$ indicating that the layer is very homogeneous. Although one would expect that this result excludes a surface depletion of Fe which was suggested by Sun \textit{et al.} \cite{Sun2012} as a source of two magnon scattering, recent XPS measurements have shown that even in layers fabricated by this method the Y/Fe ration is strongly increased at the surface. So it seems on one hand that the Fe deficiency does not change the lattice quality on a noticeable level. On the other hand in constrast to the assumption by Sun \textit{et al.} the dynamic properties of these layers are very good. A 56\,nm thick layer fabricated this way exhibited a minimum linewidth of $\mu_0 \Delta H_{\text{HWHM}}=130$\,$\mu$T at 9.6\,GHz and a damping of $\alpha\le6.5\times10^{-5}$. Fig. \ref{Hauser}. Up to now these are the lowest values for linewidth and damping ever reported for sub 100 nm films. This sample was annealed at 800$^\circ$C for 30 minutes. Another 20\,nm thick film was annealed also at 800$^\circ$C but for 3 hours. For this film 20 nm thick film the linewidth at 9.6\,GHz is larger ($\mu_0 \Delta H_{\text{HWHM}}=325$\,$\mu$T but still the damping is as low as $\alpha\le7.39\times10^{-5}$ which is also lower than other values previously reported. A higher annealing temperature resulted in a linewidth of $\mu_0 \Delta H_{\text{HWHM}} =160$\,$\mu$T at 9.6\,GHz but no damping could be determined because of overlapping spin waves in the accessible frequency range. Experience shows, however, that for the temperature range from 800$^\circ$C to 900$^\circ$C and for annealing times from 30 minutes to 4 hours the variations are merely statistical. Typically the damping is as good as $\alpha=2\times10^{-4}$ or better. Although values lower than $\alpha=10^{-4}$ can be achieved these extraordinary results cannot be repeated intentionally. The coercive field of these films for in-plane field as determined by SQUID magnetometry are below 100\,$\mu$T. $\mu_{0}M_{\text S}$ for the two films is 144\,mT (56\,nm) and 131\,mT. Both values are well below the bulk magnetisation, which apparently does not affect magnetization dynamics.

In 2017 we reported studies on YIG films deposited at room temperature by PLD but annealed in Ar~\cite{Hauser2017}. Surprisingly these experiments also yield excellent quality, although not as good as for annealing in oxygen. The best values in this set of experiments are obtained for a 65\,nm thick sample annealed for 3 hours. The linewidth at 9.6 GHz is $\mu_0 \Delta H_{\text{HWHM}}=226$\,$\mu$T and the damping is $\alpha=1.61\times10^{-4}$. For this film the rocking curve showed a FWHM of 0.014$^\circ$ and the surface roughness is as small as 0.05\,nm RMS. For a thicker film the linewidth is of similar magnitude but because of numerous additional lines caused by spin waves damping could not be determined. All these films had a coercive field below the resolution limit of the SQUID magnetometer of 30\,$\mu$T. Because the results from Chang \textit{et al.}~\cite{Chang2014} were based on sputtering in Ar and annealing in oxygen also other parameter sets were tested. In one experiment deposition in Ar with subsequent annealing in oxygen was investigated. X-ray diffraction data for this film only indicated the presence of a polycristalline YIG phase but no monocrystalline YIG film. This can be understood as oxygen needs to be incorporated into the film during growth. When in a process suitable for the deposition of YIG most of the oxygen present is replaced by Ar this effect can be expected. In another experiment we checked the necessity of a gas pressure during annealing by depositing a YIG film under suitable conditions in oxygen but annealing in vacuum. The resulting film was smooth and showed a crystalline phase different from YIG. Also no magnetic phase could be detected. So it stands to reason that during annealing an atmosphere is necessary to avoid the outdiffusion of oxygen. Interestingly this atmosphere does not need to contain oxygen. Although no further tests were performed, it is likely that also other non-reactive gases can be used.

Also in 2017 Krysztofik \textit{et al.} \cite{Krysztofik2017} reported on lift-off patterned structures which were deposited by PLD at room temperature and subsequently annealed. The layers were deposited on (001)-oriented GGG and were annealed for 30 minutes at 850$^\circ$C. The lattice constant of the strained films according to X-ray diffraction was 1.2428\,nm (larger than GGG) and the surface roughness was approx. 0.3\,nm. The microstructures fabricated by optical lithography and lift-off were as large as 0.5\,mm$\times$0.5\,mm and can almost be considered as extended films. They show a slightly smaller linewidth in FMR than the full films which may be attributed to the absence of long range spin wave propagation and reflection. At 10\,GHz the linewidth is 470\,$\mu$T and the damping is $\alpha\approx5\times10^{-4}$.

\section{Nanopatterning and lift-off}
\begin{figure}[h]%
\includegraphics*[width=\linewidth]{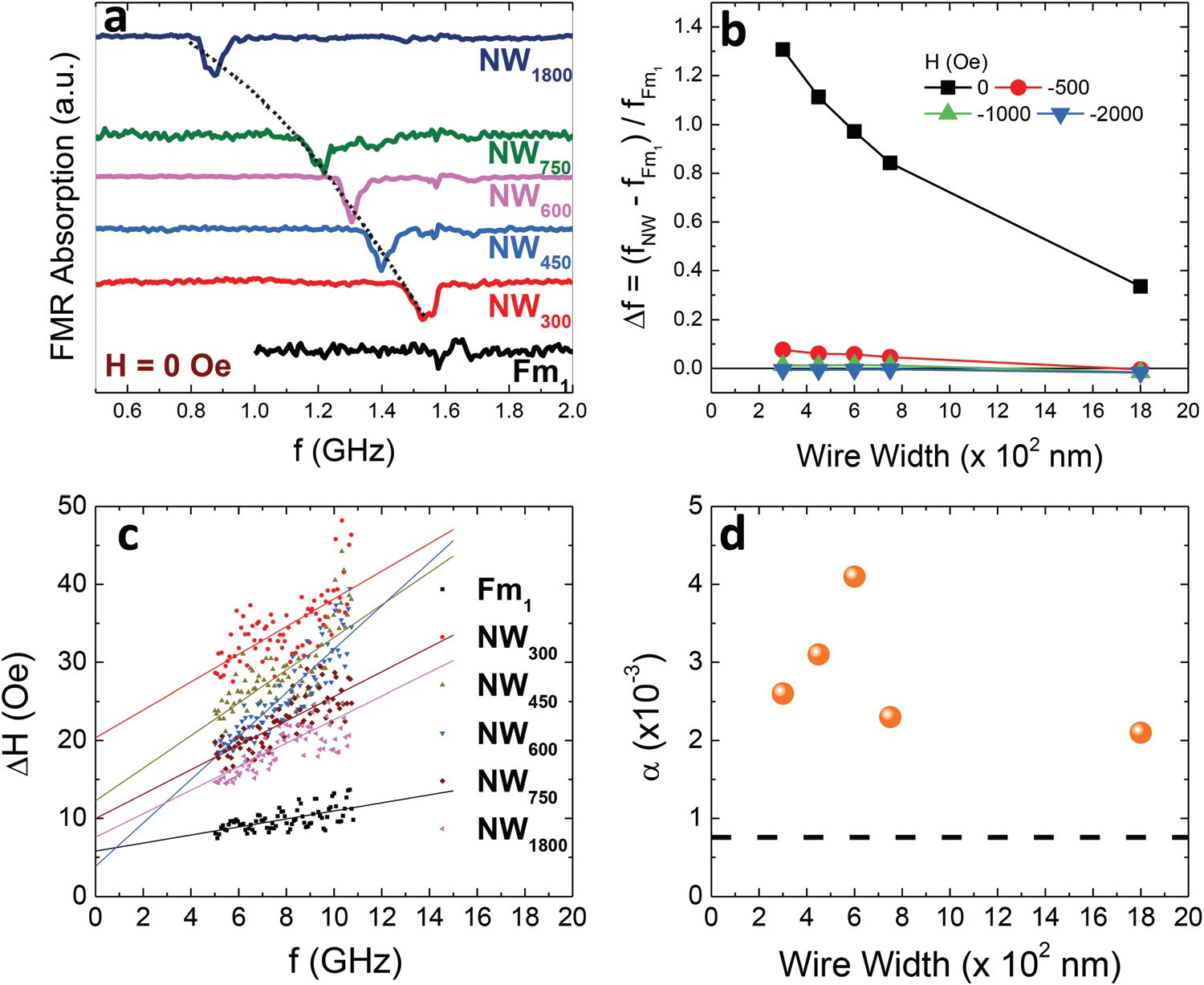}
\caption{Republished with permission of Royal Society of Chemistry (Great Britain), from Li \textit{et al.}, Nanoscale 8, 388 (2016)~\cite{Li2016}. (a) FMR 1D-spectrum of Fm1 (extended film) and NW (nanowire) series samples recorded at zero external field showing the evolution of the main FMR mode. (b) Percentage change in the resonance frequency of the main FMR mode at different external applied fields, $H$ = 0, -500, -1000, and -2000 Oe for different widths of the nanowire. (c) Resonance linewidth versus frequency of the different samples. (d) Extracted magnetic damping values for nanowires with different widths. Dashed line indicates the damping value for the continuous film.}
\label{Li6}
\end{figure}

In the following we want to describe results related to new opportunities arising from room temperature deposition of YIG. Making YIG nanostructures can be attractive for integrated magonics. For the patterning of nanostructures a number of aspects need to be considered. Although dry etching has been demonstrated in the few 100\,nm range~\cite{Hahn2014} making smaller nanostructures is not straight-forward. Ion beam etching leads to non-vertical sidewalls and redeposition and can also damage the material. The structures shown by Hahn \textit{et al.}~\cite{Hahn2014} were 700\,nm disks in a 20\,nm film, which corresponds to an aspect ration of 1:35. For higher aspect ratio it is necessary to use other techniques. It is also known that YIG can be etched by phosphoric acid. Nevertheless, wet chemical etching suffers from isotropic etch characteristics leading to under-etching and non-vertical sidewalls. Because of the isotropic etching also the aspect ratio of the fabricated structures always is smaller than 1. With the introduction of room temperature deposition there is a new path to achieving high quality nanostructures. Patterning by electron beam lithography and lift-off avoids the etch damage and can be used to obtain very small structures with almost vertical sidewalls. Lift-off processes are well established in nanopatterning. The sample surface is first covered with a resist film wich subsequently is patterned by optical or e-beam lithography. The material which is to be patterned is then deposited onto the sample. After dissolving the resist (lift-off) the material inside the resist openings remains on the sample while the material deposited on the resist floats away. Because of the limited temperature stability of typical resists the process is limited to room temperature deposition. Hence the introduction of room temperature deposition of YIG lead to several attempts of Micro or nanopatterning by lift-off with different results. In 2016 and 2017 two publications came from the group of Axel Hoffmann \cite{Li2016,Jungfleisch2017} in both of which the described technique was used. In \cite{Jungfleisch2015} YIG stripes with a width of 765 nm were fabricated but no details on linewidth or damping were given. Li \textit{et al.}~\cite{Li2016} presented arrays of nanostructures varying in size from 300\,nm to 1800\,nm respectively (Fig. \ref{Li6}). For all wire structures the damping was well above $\alpha=1\times10^{-3}$ which is much larger than for the extended film which exhibited $\alpha\approx3\times10^{-4}$. The authors do not attribute this result to damage or defects induced by the patterning process but to additional loss channels because different spin wave modes (namely edge modes) start to couple to the main resonance.

\begin{figure}[h]%
\includegraphics*[width=\linewidth]{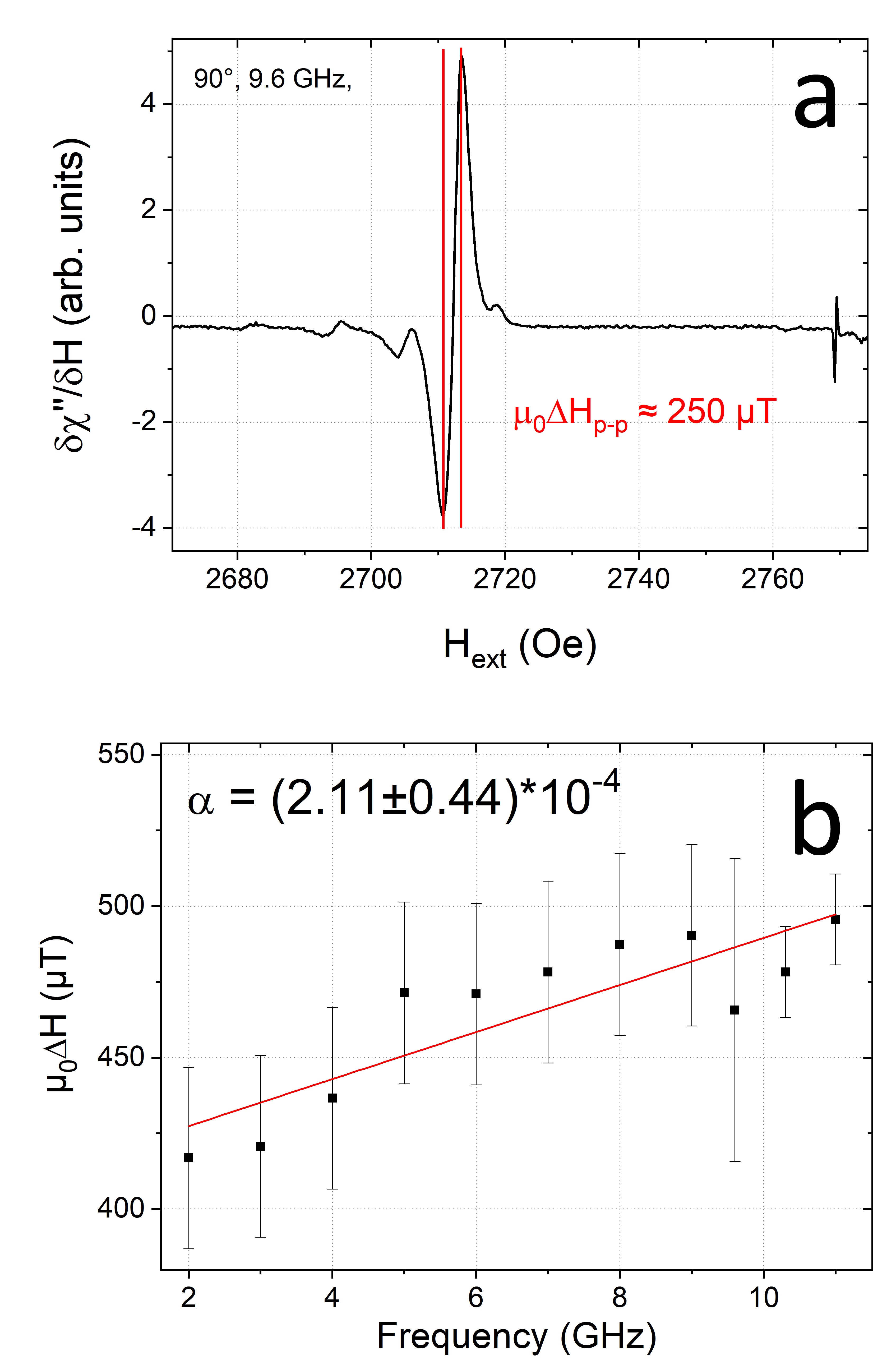}
\caption{(a) FMR measurement at 9.6 GHz on an array of 1000 nanostructures (500\,nm $\times$ 4000\,nm $\times$ 30 nm). Besides the main resonance which exhibits a linewidth of $\mu_0\Delta H_{\text{p-p}}$ of approx. 250\,$\mu$T several lines at higher and lower respective frequency appear which are due to confined spin wave modes in the structures. (b) For another array of similar dimensions the damping could be determined to $\alpha$ = (2.11$\pm$ 0.44) $\times 10^{-4}$}.
\label{Array}
\end{figure}

The fact that the patterning itself does not decrease the quality seems to be confirmed by our own results. Using electron beam lithography, room temperature PLD and lift-off as described in \cite{Hauser2016} we have fabricated arrays of 1000 nominally identical structures. The dimensions of each structure are 500\,$\times$4000\,nm with a thickness of 30\,nm. Onto the sample a coplanar waveguide was deposited in a way that the array was placed in the gap between central conductor and ground plane. Hence, FMR measurements always show an average over all structures. Despite this fact we could observe a mode with a linewidth of $\mu_0 \Delta H_{\text{HWHM}}=210$\,$\mu$T at 9.6\,GHz (Fig. \ref{Array}a). As we are averaging over 1000 structures the linewidth can well be even smaller than this. Also we were able to determine the damping for another sample of similar dimensions to $\alpha=2.11\times10^{-4}$ (Fig. \ref{Array}b) which can even be considered a very good value for a 30\,nm extended film. Apparently as suggested by Li \textit{et al.} the patterning does not influence the magnonic properties, but in our experiments the magnon frequencies might be more favourable than in \cite{Li2016}.

\begin{figure}[h]%
\includegraphics*[width=\linewidth]{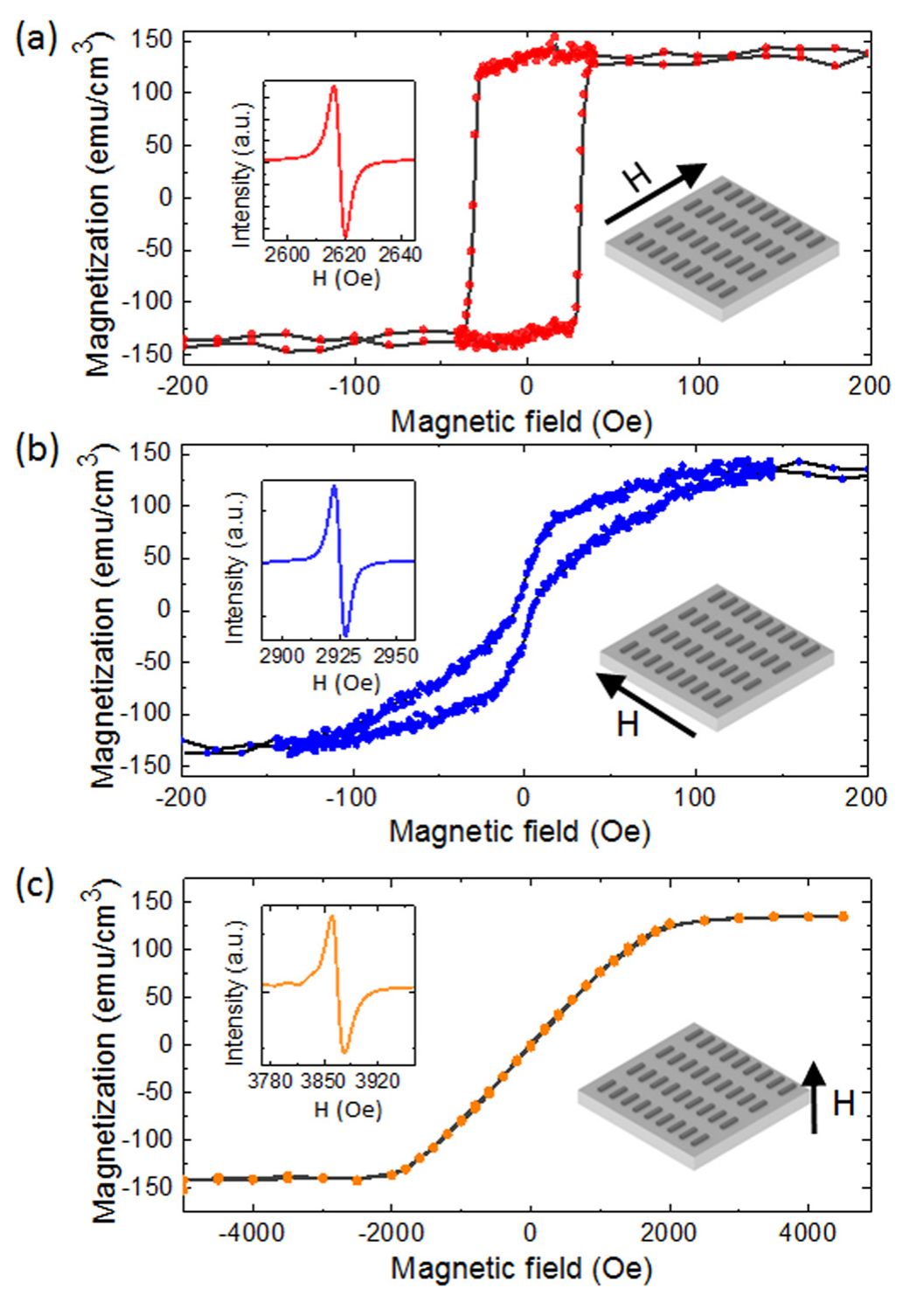}
\caption{Reproduced from Zhu \textit{et al.}, Applied Physics Letters 110(25), 252401 (2017)~\cite{Zhu2017}, with the permission of AIP Publishing. (a)-(c) Room temperature hysteresis loops of the YIG nanobars measured at different magnetic field orientations. The insets are the FMR
spectra measured at the corresponding field directions.}
\label{Zhu2}
\end{figure}

Besides the investigation of damping and spin-wave modes this kind of nanopatterning has also been used to engineer anisotropy in nanostructures. In 2017 Zhu \textit{et al.} \cite{Zhu2017} demonstrated a clear shape anisotropy in YIG films fabricated by electron beam lithography, room temperature sputtering, lift-off and subsequent annealing. The structures had a size of 3\,$\mu$m $\times$ 0.8\,$\mu$m and a thickness of 75\,nm. These rectangles showed a clear hard axis along the short side (Fig. \ref{Zhu2}) with a saturation field of more than 10\,mT. Interestingly, they also showed a largely increased coercive field along the easy axis (4\,mT compared to 0.1\,mT for the extended film) which can probably be explained by the increasing domain wall nucleation energy which is well known for ferromagnetic nanostructures.

\section{Annealing and interdiffusion}
\subsection{Annealing} As we have shown, a number of processes use room temperature deposition and annealing. Here, the annealing promotes the necessary recrystallization. When growth is performed at elevated substrate temperatures the annealing step usually can be avoided. Nevertheless, there are also reports where annealing after high temperature growth is beneficial for the realization of smooth surfaces and small FMR linewidth~\cite{Sun2012}. Sun \textit{et al.} also show that when lower temperatures are used the surface roughness increases. At the same time the authors observe a large Fe deficiency at the YIG surface which becomes even larger for lower growth and annealing temperatures.

However, annealing can also be responsible for the formation of a dead interface layer which is discussed below. YIG films can suffer from gallium or gadolinium diffusion from the GGG substrate during an annealing step as reported by Mitra \textit{et al.}~\cite{Mitra2017}. Furthermore, the existence Y$_2$O$_3$ overlayer~\cite{Cooper2017} was attributed to the annealing. Further annealing of the YIG films in vacuum at temperatures between 300$^\circ$C and 400$^\circ$C after growth and cool-down were performed by Bai \textit{et al.} \cite{Bai2019}. Although no discernible influence on composition and surface roughness was seen a number of changes took place. The damping of the films and the coercivity were considerably increased after the annealing. The authors attribute this to the introduction of oxygen vacancies which is understandable as annealing is performed in vacuum.  Also in YIG/Pt hybrid structures the interface spin transport was modified. While the ISHE decreased considerably after annealing the spin Seebeck effect increased. At the same time the authors found Y$_2$O$_3$ at the YIG surface as also observed in~\cite{Cooper2017}.

In conclusion, one can say that for both room temperature or high temperature deposition a post-deposition annealing can be mandatory or at least useful to achieve YIG films with low damping. In these cases the annealing is usually done at elevated pressures and in almost all cases in air or oxygen. For a film after deposition and cool down to the best of our knowledge no positive effects of annealing on magnetic properties were reported. On the contrary, an annealing step in vacuum even at medium temperatures seems to be detrimental for the damping of thin film YIG.

\subsection{Interdiffusion}Dead layers or interdiffusion at the YIG/GGG interface have also been reported by other groups~\cite{Mitra2017,Cooper2017,Bai2019,Suturin2018}. The dead layer was found to be either nonmagnetic, or to have a very small moment. The thickness of this layer can range from 1.2\,nm~\cite{Bai2019} to 5-7 nm~\cite{Mitra2017}. The origin of the interlayer was suggested to be either Gd~\cite{Mitra2017,Cooper2017} or Ga atoms~\cite{Suturin2018}, that diffuse from the substrate into the YIG layer due to the high growth and annealing temperature usually above 700$^\circ$C, or due to resputtering of the substrate. It should be noted that the origin of the dead layer  may differ depending on the respective deposition technique. In those cases where a dead layer exists, magnetization measurements typically result in an underestimation of $\mu_0M_{\text S}$ if the dead layer is not taken into account. It comes to mind that the fact that quite often for thin film YIG $\mu_0M_{\text S}$ often is well below the bulk value (see table \ref{Tablesummary}) might be related to dead layers. In some cases this may not be completely excluded. Nevertheless, in most cases the evidence of extremely smooth interfaces and a value of $\mu_0M_{\text S}$ that does not vary with YIG film thickness seem to contradict the presence of an interlayer with reduced magnetization.

\section{Conclusion}
In Table~\ref{Tablesummary} we have assembled the most useful parameters for the layers discussed by the various authors.

\begin{table*}[h]

\begin{center}

\caption{Summarized results of different YIG properties. For better comparison we have converted values to SI units wherever necessary. Values marked with an * have been extrapolated from data given in the respective publication or from a graph. Wherever necessary the linewidth has been converted to $\mu_{0}\Delta H_{\text{HWHM}}$.}

\begin{tabular}{@{}lp{1.5cm}p{2cm}llll@{}}

\hline

Deposition  & GGG &  Film Thickness  & $\mu_{0}M_{\text{S}}$ & $\mu_{0}\Delta H_{\text{HWHM}}$  &  Gilbert damping  & Ref.\\
technique    &          & [nm]        & [mT]                                  &  [$\mu$T]                                       & $\alpha$ $/ 10^{-4}$\,&  \\

\hline

PLD @ $650^{\circ}$C & (111)  & 20     & 210\,$\pm$\,5  &  165  @6 GHz       & 2.3                                                             &  \cite{Kelly2013}\\

PLD @ $650^{\circ}$C & (111)  & 7       & 210\,$\pm$\,5   &               & 16                                                             &  \cite{Kelly2013}\\

PLD @ $650^{\circ}$C & (111)  & 4       & 170\,$\pm$\,5    &              & 38                                                              &  \cite{Kelly2013}\\

PLD @ $650^{\circ}$C & (111)  & 20     &  210                   &                & 4                                                               &  \cite{Hahn2014}\\

PLD @ $650^{\circ}$C & (100)  & 20  $\pm$\,3   &  103\,$\pm$\,15 &                                                      & 21.69\,$\pm$\,0.69                 &  \cite{Jungfleisch2015}\\

PLD @ $650^{\circ}$C & (100)  & 75$\pm$\,10     &  166\,$\pm$\,22 &                                                   & 4.89\,$\pm$\,0.07                    &  \cite{Jungfleisch2015}\\

PLD @ $650^{\circ}$C & (100)  & 17$\pm$\,1     &  158\,$\pm$\,4 &       450 \,$\pm$\,40 @ 10\,GHz         & 7.0\,$\pm$\,0.7                      &  \cite{Onbasli2014}\\

PLD @ $650^{\circ}$C & (100)  & 34$\pm$\,1     &  170\,$\pm$\,4 &       265 \,$\pm$\,50 @ 10\,GHz         &5.8\,$\pm$\,0.5                       &  \cite{Onbasli2014}\\

PLD @ $650^{\circ}$C & (100)  & 49$\pm$\,2     &  170\,$\pm$\,4 &       240 \,$\pm$\,50 @ 10\,GHz         &4.1\,$\pm$\,0.5                       &  \cite{Onbasli2014}\\

PLD @ $650^{\circ}$C & (100)  & 64$\pm$\,2     &  167\,$\pm$\,4 &       220 \,$\pm$\,30 @ 10\,GHz         &3.8\,$\pm$\,0.2                       &  \cite{Onbasli2014}\\

PLD @ $650^{\circ}$C & (100)  & 79$\pm$\,2     &  172\,$\pm$\,4 &       150 \,$\pm$\,20 @ 10\,GHz         &2.2\,$\pm$\,0.2                       &  \cite{Onbasli2014}\\

PLD @ $650^{\circ}$C & (100)  & 92$\pm$\,2     &  172\,$\pm$\,4 &       185 \,$\pm$\,20 @ 10\,GHz         &3.4\,$\pm$\,0.3                       &  \cite{Onbasli2014}\\

PLD @ $650^{\circ}$C & (111)  & 20                 &                         &                                                                       &4.8\,$\pm$\,0.5                        &  \cite{Collet2017_AIP}\\

PLD @ $650^{\circ}$C & (111)  & 20                &      213 ($M_{\text{eff}}$)               &                                                                        &4.0                                          &  \cite{Collet2017}\\

PLD @ $750^{\circ}$C & (110)  & 100               &   $\approx$ 180(*)     &                                                                        & 1.0                                         &  \cite{Tang2016}\\

PLD @ $750^{\circ}$C & (110)  & 17               &                         &                                                                        & 7.2                                          &  \cite{Tang2016}\\

PLD @ $750^{\circ}$C & (111)  & 1220           &  178               &       78 @ 9.1 GHz                                       &                                                &  \cite{Manuilov2009}\\

PLD @ $790^{\circ}$C & (111)  & 11                &      167 ($M_{\text{eff}}$)                &       520  @ 9.5 GHz                                       &3.2\,$\pm$\,0.3                       &  \cite{Sun2012}\\

PLD @ $790^{\circ}$C & (111)  & 19                &         188 ($M_{\text{eff}}$)             &       294  @ 9.5 GHz                                       &2.3\,$\pm$\,0.1                       &  \cite{Sun2012}\\

PLD @ $800^{\circ}$C & (111)  & 40                &  144             &                                                                         &3.5\,$\pm$\,0.3                       &  \cite{Qin2018}\\

PLD @ $825^{\circ}$C & (111)  & 23                &      160                &      173-268 @ 9.5 GHz                                  &1.8                                          &  \cite{Howe2015}\\

PLD @ $700-850^{\circ}$C & (111)  & 1000             &                      &      50 @ 9.5 GHz                                  &                                          &  \cite{Dorsey1993}\\

PLD @ RT                & (111)     & 20     & 130.7\,$\pm$\,0.25  & 349 \,$\pm$\,10 @ 9.6 GHz   &  0.739\,$\pm$\,0.14           &  \cite{Hauser2016} \\

PLD @ RT                 & (111)    & 56     & 144\,$\pm$\,0.25\,   & 130\,$\pm$\,5 @ 9.6 GHz     & 0.615\,$\pm$\,0.15            &  \cite{Hauser2016}\\

PLD @ RT                 & (111)    & 65     &                                     & 226  @ 9.6 GHz                  & 1.61\,$\pm$\,0.25                                 &  \cite{Hauser2017}\\

PLD @ RT                 & (111)    & 65     &                                     & 511  @ 9.6 GHz                  &                                               &  \cite{Hauser2017}\\

PLD @ RT                 & (111)    & 65     &                                     & 709  @ 9.6 GHz                  &                                               &  \cite{Hauser2017}\\

PLD @ RT                 & (111)    & 130     &     179.6                                & 603  @ 9.6 GHz                  &                                               &  \cite{Hauser2017}\\

PLD @ RT                 & (111)    & 130     &                                     & 229  @ 9.6 GHz                  &                                               &  \cite{Hauser2017}\\

PLD @ RT                  & (001)    & 70     &148                  &     470  @ 10 GHz        &   5.0\,$\pm$\,0.1, 5.5\,$\pm$\,0.1                                       &  \cite{Krysztofik2017}\\

Sputter @ RT                & (111)  & 19            &  100\,$\pm$\,12   &                                                             &8\,$\pm$\,2                       & \cite{Talalaevskij2017}\\

Sputter @ RT                & (111)  & 29            &  140\,$\pm$\,20   &                                                             &5.8\,$\pm$\,0.7                       & \cite{Talalaevskij2017}\\

Sputter @ RT                & (111)  & 38            &  150\,$\pm$\,10   &                                                             &2.6\,$\pm$\,0.3                       & \cite{Talalaevskij2017}\\

Sputter @ RT                & (111)  & 49            &  160\,$\pm$\,10   &             400 @10 GHz                    &2.4\,$\pm$\,0.3                       & \cite{Talalaevskij2017}\\

Sputter @ RT                & (111)  & 40            &  153\,$\pm$\,0.4    &                                                              &2.77\,$\pm$\,0.49                       &  \cite{Jungfleisch2015}\\

Sputter @ RT                & (111)  & 75            &                               &         346 @ 9.868 GHz                                                     &                      &  \cite{Zhu2017}\\

Sputter @ RT                & (111)  & 40            &  163   &                                                                                          &2.93, 3.53, 7.56                  &  \cite{Li2016}\\

Sputter @ RT                & (111)  & 22                &         177             &       641  @ 16.5 GHz                                       &0.858\,$\pm$\,0.21                       &  \cite{Chang2014}\\

Sputter @ RT                & (111)  & 23.4         &  175 - 192 ($M_{\text{eff}}$)              &   850 - 1200 @ 16.5 GHz      & 0.85- 0.94\,$\pm$\,0.2             &  \cite{Chang2017}\\

Sputter @ RT                & (111)  & 96                &          129 \,$\pm$\,5           &       329  @  9.45 GHz                   &7.0\,$\pm$\,1.0                       &  \cite{Lustikova2014}\\

Off-axis Sputter @ $750^{\circ}$C               & (111)  & 30                &                &       238  @  9.6 GHz                                  &                                       & \cite{Yang2018}\\

\hline

\end{tabular}

\label{Tablesummary}

\end{center}

\end{table*}

The results presented here clearly show that for ultrathin YIG films other methods than LPE may be more suitable. PLD at elevated temperatures yields layers down to 20\,nm thickness with a damping in the lower $10^{-4}$ range exhibiting linewidths which may be smaller than 200\,$\mu$T at 9.5\,GHz. Room temperature deposition and annealing either by sputtering or by PLD yields even lower damping, however the linewidth with sputtering is slightly higher than the best values for PLD at elevated substrate temperature. The best values in terms of both, damping and linewidth have been achieved by room temperature PLD and annealing. In all cases (as also in high quality LPE thick films) GGG was used as a substrate because it presents almost perfect lattice matching to YIG. Interestingly within certain limits, the saturation magnetization and also the lattice constant of the films do not seem to influence the dynamic properties of the films. In total, the best results are obtained by room temperature methods which as a bonus also offer the possibility to fabricate YIG nanostructures by lift-off. Although it is difficult to say what parameters determine whether the quality of a grown layer is high or low one can at least state that all films with low damping also show low surface roughness which is consistent with the assumption of two magnon scattering as one of the critical processes influencing linewidth and damping. Also, when measured, PLD grown films showed an Fe deficiency at the surface, however, this does not to be critical as it was also observed in the very best layers.

\begin{acknowledgements}
This work was funded by the Deutsche Forschungsgemeinschaft in the SFB 762.
\end{acknowledgements}

%

%


\clearpage
\begin{thebibliography}{44}%
\makeatletter
\providecommand \@ifxundefined [1]{%
 \@ifx{#1\undefined}
}%
\providecommand \@ifnum [1]{%
 \ifnum #1\expandafter \@firstoftwo
 \else \expandafter \@secondoftwo
 \fi
}%
\providecommand \@ifx [1]{%
 \ifx #1\expandafter \@firstoftwo
 \else \expandafter \@secondoftwo
 \fi
}%
\providecommand \natexlab [1]{#1}%
\providecommand \enquote  [1]{``#1''}%
\providecommand \bibnamefont  [1]{#1}%
\providecommand \bibfnamefont [1]{#1}%
\providecommand \citenamefont [1]{#1}%
\providecommand \href@noop [0]{\@secondoftwo}%
\providecommand \href [0]{\begingroup \@sanitize@url \@href}%
\providecommand \@href[1]{\@@startlink{#1}\@@href}%
\providecommand \@@href[1]{\endgroup#1\@@endlink}%
\providecommand \@sanitize@url [0]{\catcode `\\12\catcode `\$12\catcode
  `\&12\catcode `\#12\catcode `\^12\catcode `\_12\catcode `\%12\relax}%
\providecommand \@@startlink[1]{}%
\providecommand \@@endlink[0]{}%
\providecommand \url  [0]{\begingroup\@sanitize@url \@url }%
\providecommand \@url [1]{\endgroup\@href {#1}{\urlprefix }}%
\providecommand \urlprefix  [0]{URL }%
\providecommand \Eprint [0]{\href }%
\providecommand \doibase [0]{https://doi.org/}%
\providecommand \selectlanguage [0]{\@gobble}%
\providecommand \bibinfo  [0]{\@secondoftwo}%
\providecommand \bibfield  [0]{\@secondoftwo}%
\providecommand \translation [1]{[#1]}%
\providecommand \BibitemOpen [0]{}%
\providecommand \bibitemStop [0]{}%
\providecommand \bibitemNoStop [0]{.\EOS\space}%
\providecommand \EOS [0]{\spacefactor3000\relax}%
\providecommand \BibitemShut  [1]{\csname bibitem#1\endcsname}%
\let\auto@bib@innerbib\@empty
\bibitem [{\citenamefont {Kruglyak}\ \emph {et~al.}(2010)\citenamefont
  {Kruglyak}, \citenamefont {Demokritov},\ and\ \citenamefont
  {Grundler}}]{Kruglyak2010}%
  \BibitemOpen
  \bibfield  {author} {\bibinfo {author} {\bibfnamefont {V.~V.}\ \bibnamefont
  {Kruglyak}}, \bibinfo {author} {\bibfnamefont {S.~O.}\ \bibnamefont
  {Demokritov}},\ and\ \bibinfo {author} {\bibfnamefont {D.}~\bibnamefont
  {Grundler}},\ }\bibfield  {title} {\bibinfo {title} {Magnonics},\ }\href
  {https://doi.org/10.1088/0022-3727/43/26/264001} {\bibfield  {journal}
  {\bibinfo  {journal} {Journal of Physics D: Applied Physics}\ }\textbf
  {\bibinfo {volume} {43}},\ \bibinfo {pages} {264001} (\bibinfo {year}
  {2010})}\BibitemShut {NoStop}%
\bibitem [{\citenamefont {Khitun}(2012)}]{Khitun2012}%
  \BibitemOpen
  \bibfield  {author} {\bibinfo {author} {\bibfnamefont {A.}~\bibnamefont
  {Khitun}},\ }\bibfield  {title} {\bibinfo {title} {Multi-frequency magnonic
  logic circuits for parallel data processing},\ }\href
  {https://doi.org/10.1063/1.3689011} {\bibfield  {journal} {\bibinfo
  {journal} {Journal of Applied Physics}\ }\textbf {\bibinfo {volume} {111}},\
  \bibinfo {pages} {054307} (\bibinfo {year} {2012})},\ \Eprint
  {https://arxiv.org/abs/https://doi.org/10.1063/1.3689011}
  {https://doi.org/10.1063/1.3689011} \BibitemShut {NoStop}%
\bibitem [{\citenamefont {Chumak}\ \emph {et~al.}(2015)\citenamefont {Chumak},
  \citenamefont {Vasyuchka}, \citenamefont {Serga},\ and\ \citenamefont
  {Hillebrands}}]{Chumak2015}%
  \BibitemOpen
  \bibfield  {author} {\bibinfo {author} {\bibfnamefont {A.~V.}\ \bibnamefont
  {Chumak}}, \bibinfo {author} {\bibfnamefont {V.~I.}\ \bibnamefont
  {Vasyuchka}}, \bibinfo {author} {\bibfnamefont {A.~A.}\ \bibnamefont
  {Serga}},\ and\ \bibinfo {author} {\bibfnamefont {B.}~\bibnamefont
  {Hillebrands}},\ }\bibfield  {title} {\bibinfo {title} {Magnon spintronics},\
  }\href {https://doi.org/10.1038/nphys3347} {\bibfield  {journal} {\bibinfo
  {journal} {Nature Physics}\ }\textbf {\bibinfo {volume} {11}},\ \bibinfo
  {pages} {453} (\bibinfo {year} {2015})}\BibitemShut {NoStop}%
\bibitem [{\citenamefont {Lenk}\ \emph {et~al.}(2011)\citenamefont {Lenk},
  \citenamefont {Ulrichs}, \citenamefont {Garbs},\ and\ \citenamefont
  {M{\"u}nzenberg}}]{Lenk2011}%
  \BibitemOpen
  \bibfield  {author} {\bibinfo {author} {\bibfnamefont {B.}~\bibnamefont
  {Lenk}}, \bibinfo {author} {\bibfnamefont {H.}~\bibnamefont {Ulrichs}},
  \bibinfo {author} {\bibfnamefont {F.}~\bibnamefont {Garbs}},\ and\ \bibinfo
  {author} {\bibfnamefont {M.}~\bibnamefont {M{\"u}nzenberg}},\ }\bibfield
  {title} {\bibinfo {title} {The building blocks of magnonics},\ }\href
  {https://doi.org/https://doi.org/10.1016/j.physrep.2011.06.003} {\bibfield
  {journal} {\bibinfo  {journal} {Physics Reports}\ }\textbf {\bibinfo {volume}
  {507}},\ \bibinfo {pages} {107 } (\bibinfo {year} {2011})}\BibitemShut
  {NoStop}%
\bibitem [{\citenamefont {Khitun}\ \emph {et~al.}(2010)\citenamefont {Khitun},
  \citenamefont {Bao},\ and\ \citenamefont {Wang}}]{Khitun2010}%
  \BibitemOpen
  \bibfield  {author} {\bibinfo {author} {\bibfnamefont {A.}~\bibnamefont
  {Khitun}}, \bibinfo {author} {\bibfnamefont {M.}~\bibnamefont {Bao}},\ and\
  \bibinfo {author} {\bibfnamefont {K.~L.}\ \bibnamefont {Wang}},\ }\bibfield
  {title} {\bibinfo {title} {Magnonic logic circuits},\ }\href
  {https://doi.org/10.1088/0022-3727/43/26/264005} {\bibfield  {journal}
  {\bibinfo  {journal} {Journal of Physics D: Applied Physics}\ }\textbf
  {\bibinfo {volume} {43}},\ \bibinfo {pages} {264005} (\bibinfo {year}
  {2010})}\BibitemShut {NoStop}%
\bibitem [{\citenamefont {Fischer}\ \emph {et~al.}(2017)\citenamefont
  {Fischer}, \citenamefont {Kewenig}, \citenamefont {Bozhko}, \citenamefont
  {Serga}, \citenamefont {Syvorotka}, \citenamefont {Ciubotaru}, \citenamefont
  {Adelmann}, \citenamefont {Hillebrands},\ and\ \citenamefont
  {Chumak}}]{Fischer2017}%
  \BibitemOpen
  \bibfield  {author} {\bibinfo {author} {\bibfnamefont {T.}~\bibnamefont
  {Fischer}}, \bibinfo {author} {\bibfnamefont {M.}~\bibnamefont {Kewenig}},
  \bibinfo {author} {\bibfnamefont {D.~A.}\ \bibnamefont {Bozhko}}, \bibinfo
  {author} {\bibfnamefont {A.~A.}\ \bibnamefont {Serga}}, \bibinfo {author}
  {\bibfnamefont {I.~I.}\ \bibnamefont {Syvorotka}}, \bibinfo {author}
  {\bibfnamefont {F.}~\bibnamefont {Ciubotaru}}, \bibinfo {author}
  {\bibfnamefont {C.}~\bibnamefont {Adelmann}}, \bibinfo {author}
  {\bibfnamefont {B.}~\bibnamefont {Hillebrands}},\ and\ \bibinfo {author}
  {\bibfnamefont {A.~V.}\ \bibnamefont {Chumak}},\ }\bibfield  {title}
  {\bibinfo {title} {Experimental prototype of a spin-wave majority gate},\
  }\href {https://doi.org/10.1063/1.4979840} {\bibfield  {journal} {\bibinfo
  {journal} {Applied Physics Letters}\ }\textbf {\bibinfo {volume} {110}},\
  \bibinfo {pages} {152401} (\bibinfo {year} {2017})},\ \Eprint
  {https://arxiv.org/abs/https://doi.org/10.1063/1.4979840}
  {https://doi.org/10.1063/1.4979840} \BibitemShut {NoStop}%
\bibitem [{\citenamefont {Krawczyk}\ and\ \citenamefont
  {Grundler}(2014)}]{Krawczyk2014}%
  \BibitemOpen
  \bibfield  {author} {\bibinfo {author} {\bibfnamefont {M.}~\bibnamefont
  {Krawczyk}}\ and\ \bibinfo {author} {\bibfnamefont {D.}~\bibnamefont
  {Grundler}},\ }\bibfield  {title} {\bibinfo {title} {Review and prospects of
  magnonic crystals and devices with reprogrammable band structure},\ }\href
  {https://doi.org/10.1088/0953-8984/26/12/123202} {\bibfield  {journal}
  {\bibinfo  {journal} {Journal of Physics: Condensed Matter}\ }\textbf
  {\bibinfo {volume} {26}},\ \bibinfo {pages} {123202} (\bibinfo {year}
  {2014})}\BibitemShut {NoStop}%
\bibitem [{\citenamefont {Chumak}\ \emph {et~al.}(2014)\citenamefont {Chumak},
  \citenamefont {Serga},\ and\ \citenamefont {Hillebrands}}]{Chumak2014}%
  \BibitemOpen
  \bibfield  {author} {\bibinfo {author} {\bibfnamefont {A.~V.}\ \bibnamefont
  {Chumak}}, \bibinfo {author} {\bibfnamefont {A.~A.}\ \bibnamefont {Serga}},\
  and\ \bibinfo {author} {\bibfnamefont {B.}~\bibnamefont {Hillebrands}},\
  }\bibfield  {title} {\bibinfo {title} {Magnon transistor for all-magnon data
  processing},\ }\href {https://doi.org/10.1038/ncomms5700} {\bibfield
  {journal} {\bibinfo  {journal} {Nature Communications}\ }\textbf {\bibinfo
  {volume} {5}},\ \bibinfo {pages} {4700} (\bibinfo {year} {2014})}\BibitemShut
  {NoStop}%
\bibitem [{\citenamefont {Yu}\ \emph {et~al.}(2016)\citenamefont {Yu},
  \citenamefont {d'Allivy Kelly}, \citenamefont {Cros}, \citenamefont
  {Bernard}, \citenamefont {Bortolotti}, \citenamefont {Anane}, \citenamefont
  {Brandl}, \citenamefont {Heimbach},\ and\ \citenamefont {Grundler}}]{Yu2016}%
  \BibitemOpen
  \bibfield  {author} {\bibinfo {author} {\bibfnamefont {H.}~\bibnamefont
  {Yu}}, \bibinfo {author} {\bibfnamefont {O.}~\bibnamefont {d'Allivy Kelly}},
  \bibinfo {author} {\bibfnamefont {V.}~\bibnamefont {Cros}}, \bibinfo {author}
  {\bibfnamefont {R.}~\bibnamefont {Bernard}}, \bibinfo {author} {\bibfnamefont
  {P.}~\bibnamefont {Bortolotti}}, \bibinfo {author} {\bibfnamefont
  {A.}~\bibnamefont {Anane}}, \bibinfo {author} {\bibfnamefont
  {F.}~\bibnamefont {Brandl}}, \bibinfo {author} {\bibfnamefont
  {F.}~\bibnamefont {Heimbach}},\ and\ \bibinfo {author} {\bibfnamefont
  {D.}~\bibnamefont {Grundler}},\ }\bibfield  {title} {\bibinfo {title}
  {Approaching soft x-ray wavelengths in nanomagnet-based microwave
  technology},\ }\href {https://doi.org/10.1038/ncomms11255} {\bibfield
  {journal} {\bibinfo  {journal} {Nature Communications}\ }\textbf {\bibinfo
  {volume} {7}},\ \bibinfo {pages} {11255} (\bibinfo {year}
  {2016})}\BibitemShut {NoStop}%
\bibitem [{\citenamefont {Urazhdin}\ \emph {et~al.}(2014)\citenamefont
  {Urazhdin}, \citenamefont {Demidov}, \citenamefont {Ulrichs}, \citenamefont
  {Kendziorczyk}, \citenamefont {Kuhn}, \citenamefont {Leuthold}, \citenamefont
  {Wilde},\ and\ \citenamefont {Demokritov}}]{Urazhdin2014}%
  \BibitemOpen
  \bibfield  {author} {\bibinfo {author} {\bibfnamefont {S.}~\bibnamefont
  {Urazhdin}}, \bibinfo {author} {\bibfnamefont {V.~E.}\ \bibnamefont
  {Demidov}}, \bibinfo {author} {\bibfnamefont {H.}~\bibnamefont {Ulrichs}},
  \bibinfo {author} {\bibfnamefont {T.}~\bibnamefont {Kendziorczyk}}, \bibinfo
  {author} {\bibfnamefont {T.}~\bibnamefont {Kuhn}}, \bibinfo {author}
  {\bibfnamefont {J.}~\bibnamefont {Leuthold}}, \bibinfo {author}
  {\bibfnamefont {G.}~\bibnamefont {Wilde}},\ and\ \bibinfo {author}
  {\bibfnamefont {S.~O.}\ \bibnamefont {Demokritov}},\ }\bibfield  {title}
  {\bibinfo {title} {Nanomagnonic devices based on the spin-transfer torque},\
  }\href {https://doi.org/10.1038/nnano.2014.88} {\bibfield  {journal}
  {\bibinfo  {journal} {Nature Nanotechnology}\ }\textbf {\bibinfo {volume}
  {9}},\ \bibinfo {pages} {509 EP } (\bibinfo {year} {2014})}\BibitemShut
  {NoStop}%
\bibitem [{\citenamefont {{Tabuchi}}\ \emph {et~al.}(2015)\citenamefont
  {{Tabuchi}}, \citenamefont {{Ishino}}, \citenamefont {{Noguchi}},
  \citenamefont {{Ishikawa}}, \citenamefont {{Yamazaki}}, \citenamefont
  {{Usami}},\ and\ \citenamefont {{Nakamura}}}]{Tabuchi2015}%
  \BibitemOpen
  \bibfield  {author} {\bibinfo {author} {\bibfnamefont {Y.}~\bibnamefont
  {{Tabuchi}}}, \bibinfo {author} {\bibfnamefont {S.}~\bibnamefont {{Ishino}}},
  \bibinfo {author} {\bibfnamefont {A.}~\bibnamefont {{Noguchi}}}, \bibinfo
  {author} {\bibfnamefont {T.}~\bibnamefont {{Ishikawa}}}, \bibinfo {author}
  {\bibfnamefont {R.}~\bibnamefont {{Yamazaki}}}, \bibinfo {author}
  {\bibfnamefont {K.}~\bibnamefont {{Usami}}},\ and\ \bibinfo {author}
  {\bibfnamefont {Y.}~\bibnamefont {{Nakamura}}},\ }\bibfield  {title}
  {\bibinfo {title} {{Coherent coupling between a ferromagnetic magnon and a
  superconducting qubit}},\ }\href {https://doi.org/10.1126/science.aaa3693}
  {\bibfield  {journal} {\bibinfo  {journal} {Science}\ }\textbf {\bibinfo
  {volume} {349}},\ \bibinfo {pages} {405} (\bibinfo {year} {2015})},\ \Eprint
  {https://arxiv.org/abs/1410.3781} {arXiv:1410.3781 [quant-ph]} \BibitemShut
  {NoStop}%
\bibitem [{\citenamefont {Glass}\ and\ \citenamefont
  {Elliot}(1976)}]{Glass1976}%
  \BibitemOpen
  \bibfield  {author} {\bibinfo {author} {\bibfnamefont {H.}~\bibnamefont
  {Glass}}\ and\ \bibinfo {author} {\bibfnamefont {M.}~\bibnamefont {Elliot}},\
  }\bibfield  {title} {\bibinfo {title} {Attainment of the intrinsic fmr
  linewidth in yttrium iron garnet films grown by liquid phase epitaxy},\
  }\href {https://doi.org/https://doi.org/10.1016/0022-0248(76)90141-X}
  {\bibfield  {journal} {\bibinfo  {journal} {Journal of Crystal Growth}\
  }\textbf {\bibinfo {volume} {34}},\ \bibinfo {pages} {285 } (\bibinfo {year}
  {1976})}\BibitemShut {NoStop}%
\bibitem [{\citenamefont {Dorsey}\ \emph {et~al.}(1993)\citenamefont {Dorsey},
  \citenamefont {Bushnell}, \citenamefont {Seed},\ and\ \citenamefont
  {Vittoria}}]{Dorsey1993}%
  \BibitemOpen
  \bibfield  {author} {\bibinfo {author} {\bibfnamefont {P.~C.}\ \bibnamefont
  {Dorsey}}, \bibinfo {author} {\bibfnamefont {S.~E.}\ \bibnamefont
  {Bushnell}}, \bibinfo {author} {\bibfnamefont {R.~G.}\ \bibnamefont {Seed}},\
  and\ \bibinfo {author} {\bibfnamefont {C.}~\bibnamefont {Vittoria}},\
  }\bibfield  {title} {\bibinfo {title} {Epitaxial yttrium iron garnet films
  grown by pulsed laser deposition},\ }\href {https://doi.org/10.1063/1.354927}
  {\bibfield  {journal} {\bibinfo  {journal} {Journal of Applied Physics}\
  }\textbf {\bibinfo {volume} {74}},\ \bibinfo {pages} {1242} (\bibinfo {year}
  {1993})},\ \Eprint {https://arxiv.org/abs/https://doi.org/10.1063/1.354927}
  {https://doi.org/10.1063/1.354927} \BibitemShut {NoStop}%
\bibitem [{\citenamefont {Hauser}\ \emph {et~al.}(2016)\citenamefont {Hauser},
  \citenamefont {Richter}, \citenamefont {Homonnay}, \citenamefont
  {Eisenschmidt}, \citenamefont {Qaid}, \citenamefont {Deniz}, \citenamefont
  {Hesse}, \citenamefont {Sawicki}, \citenamefont {Ebbinghaus},\ and\
  \citenamefont {Schmidt}}]{Hauser2016}%
  \BibitemOpen
  \bibfield  {author} {\bibinfo {author} {\bibfnamefont {C.}~\bibnamefont
  {Hauser}}, \bibinfo {author} {\bibfnamefont {T.}~\bibnamefont {Richter}},
  \bibinfo {author} {\bibfnamefont {N.}~\bibnamefont {Homonnay}}, \bibinfo
  {author} {\bibfnamefont {C.}~\bibnamefont {Eisenschmidt}}, \bibinfo {author}
  {\bibfnamefont {M.}~\bibnamefont {Qaid}}, \bibinfo {author} {\bibfnamefont
  {H.}~\bibnamefont {Deniz}}, \bibinfo {author} {\bibfnamefont
  {D.}~\bibnamefont {Hesse}}, \bibinfo {author} {\bibfnamefont
  {M.}~\bibnamefont {Sawicki}}, \bibinfo {author} {\bibfnamefont {S.~G.}\
  \bibnamefont {Ebbinghaus}},\ and\ \bibinfo {author} {\bibfnamefont
  {G.}~\bibnamefont {Schmidt}},\ }\bibfield  {title} {\bibinfo {title} {Yttrium
  iron garnet thin films with very low damping obtained by recrystallization of
  amorphous material},\ }\href {https://doi.org/10.1038/srep20827} {\bibfield
  {journal} {\bibinfo  {journal} {Scientific Reports}\ }\textbf {\bibinfo
  {volume} {6}},\ \bibinfo {pages} {20827} (\bibinfo {year}
  {2016})}\BibitemShut {NoStop}%
\bibitem [{\citenamefont {Chang}\ \emph {et~al.}(2014)\citenamefont {Chang},
  \citenamefont {Li}, \citenamefont {Zhang}, \citenamefont {Liu}, \citenamefont
  {Hoffmann}, \citenamefont {Deng},\ and\ \citenamefont {Wu}}]{Chang2014}%
  \BibitemOpen
  \bibfield  {author} {\bibinfo {author} {\bibfnamefont {H.}~\bibnamefont
  {Chang}}, \bibinfo {author} {\bibfnamefont {P.}~\bibnamefont {Li}}, \bibinfo
  {author} {\bibfnamefont {W.}~\bibnamefont {Zhang}}, \bibinfo {author}
  {\bibfnamefont {T.}~\bibnamefont {Liu}}, \bibinfo {author} {\bibfnamefont
  {A.}~\bibnamefont {Hoffmann}}, \bibinfo {author} {\bibfnamefont
  {L.}~\bibnamefont {Deng}},\ and\ \bibinfo {author} {\bibfnamefont
  {M.}~\bibnamefont {Wu}},\ }\bibfield  {title} {\bibinfo {title}
  {Nanometer-thick yttrium iron garnet films with extremely low damping},\
  }\bibfield  {booktitle} {\emph {\bibinfo {booktitle} {IEEE Magnetics
  Letters}},\ }\href {https://doi.org/10.1109/LMAG.2014.2350958} {\bibfield
  {journal} {\bibinfo  {journal} {IEEE Magnetics Letters}\ }\textbf {\bibinfo
  {volume} {5}},\ \bibinfo {pages} {1} (\bibinfo {year} {2014})}\BibitemShut
  {NoStop}%
\bibitem [{\citenamefont {Yu}\ \emph {et~al.}(2014)\citenamefont {Yu},
  \citenamefont {d'Allivy Kelly}, \citenamefont {Cros}, \citenamefont
  {Bernard}, \citenamefont {Bortolotti}, \citenamefont {Anane}, \citenamefont
  {Brandl}, \citenamefont {Huber}, \citenamefont {Stasinopoulos},\ and\
  \citenamefont {Grundler}}]{Yu2014}%
  \BibitemOpen
  \bibfield  {author} {\bibinfo {author} {\bibfnamefont {H.}~\bibnamefont
  {Yu}}, \bibinfo {author} {\bibfnamefont {O.}~\bibnamefont {d'Allivy Kelly}},
  \bibinfo {author} {\bibfnamefont {V.}~\bibnamefont {Cros}}, \bibinfo {author}
  {\bibfnamefont {R.}~\bibnamefont {Bernard}}, \bibinfo {author} {\bibfnamefont
  {P.}~\bibnamefont {Bortolotti}}, \bibinfo {author} {\bibfnamefont
  {A.}~\bibnamefont {Anane}}, \bibinfo {author} {\bibfnamefont
  {F.}~\bibnamefont {Brandl}}, \bibinfo {author} {\bibfnamefont
  {R.}~\bibnamefont {Huber}}, \bibinfo {author} {\bibfnamefont
  {I.}~\bibnamefont {Stasinopoulos}},\ and\ \bibinfo {author} {\bibfnamefont
  {D.}~\bibnamefont {Grundler}},\ }\bibfield  {title} {\bibinfo {title}
  {Magnetic thin-film insulator with ultra-low spin wave damping for coherent
  nanomagnonics},\ }\href {https://doi.org/10.1038/srep06848} {\bibfield
  {journal} {\bibinfo  {journal} {Scientific Reports}\ }\textbf {\bibinfo
  {volume} {4}},\ \bibinfo {pages} {6848} (\bibinfo {year} {2014})}\BibitemShut
  {NoStop}%
\bibitem [{\citenamefont {{Talalaevskij}}\ \emph {et~al.}(2017)\citenamefont
  {{Talalaevskij}}, \citenamefont {{Decker}}, \citenamefont {{Stigloher}},
  \citenamefont {{Mitra}}, \citenamefont {{K{\"o}rner}}, \citenamefont
  {{Cespedes}}, \citenamefont {{Back}},\ and\ \citenamefont
  {{Hickey}}}]{Talalaevskij2017}%
  \BibitemOpen
  \bibfield  {author} {\bibinfo {author} {\bibfnamefont {A.}~\bibnamefont
  {{Talalaevskij}}}, \bibinfo {author} {\bibfnamefont {M.}~\bibnamefont
  {{Decker}}}, \bibinfo {author} {\bibfnamefont {J.}~\bibnamefont
  {{Stigloher}}}, \bibinfo {author} {\bibfnamefont {A.}~\bibnamefont
  {{Mitra}}}, \bibinfo {author} {\bibfnamefont {H.~S.}\ \bibnamefont
  {{K{\"o}rner}}}, \bibinfo {author} {\bibfnamefont {O.}~\bibnamefont
  {{Cespedes}}}, \bibinfo {author} {\bibfnamefont {C.~H.}\ \bibnamefont
  {{Back}}},\ and\ \bibinfo {author} {\bibfnamefont {B.~J.}\ \bibnamefont
  {{Hickey}}},\ }\bibfield  {title} {\bibinfo {title} {{Magnetic properties of
  spin waves in thin yttrium iron garnet films}},\ }\href
  {https://doi.org/10.1103/PhysRevB.95.064409} {\bibfield  {journal} {\bibinfo
  {journal} {Physical Review B}\ }\textbf {\bibinfo {volume} {95}},\ \bibinfo
  {eid} {064409} (\bibinfo {year} {2017})}\BibitemShut {NoStop}%
\bibitem [{\citenamefont {Qin}\ \emph {et~al.}(2018)\citenamefont {Qin},
  \citenamefont {H\"am\"al\"ainen}, \citenamefont {Arjas}, \citenamefont
  {Witteveen},\ and\ \citenamefont {van Dijken}}]{Qin2018}%
  \BibitemOpen
  \bibfield  {author} {\bibinfo {author} {\bibfnamefont {H.}~\bibnamefont
  {Qin}}, \bibinfo {author} {\bibfnamefont {S.~J.}\ \bibnamefont
  {H\"am\"al\"ainen}}, \bibinfo {author} {\bibfnamefont {K.}~\bibnamefont
  {Arjas}}, \bibinfo {author} {\bibfnamefont {J.}~\bibnamefont {Witteveen}},\
  and\ \bibinfo {author} {\bibfnamefont {S.}~\bibnamefont {van Dijken}},\
  }\bibfield  {title} {\bibinfo {title} {Propagating spin waves in
  nanometer-thick yttrium iron garnet films: Dependence on wave vector,
  magnetic field strength, and angle},\ }\href
  {https://doi.org/10.1103/PhysRevB.98.224422} {\bibfield  {journal} {\bibinfo
  {journal} {Phys. Rev. B}\ }\textbf {\bibinfo {volume} {98}},\ \bibinfo
  {pages} {224422} (\bibinfo {year} {2018})}\BibitemShut {NoStop}%
\bibitem [{\citenamefont {Collet}\ \emph
  {et~al.}(2017{\natexlab{a}})\citenamefont {Collet}, \citenamefont {Gladii},
  \citenamefont {Evelt}, \citenamefont {Bessonov}, \citenamefont {Soumah},
  \citenamefont {Bortolotti}, \citenamefont {Demokritov}, \citenamefont
  {Henry}, \citenamefont {Cros}, \citenamefont {Bailleul}, \citenamefont
  {Demidov},\ and\ \citenamefont {Anane}}]{Collet2017}%
  \BibitemOpen
  \bibfield  {author} {\bibinfo {author} {\bibfnamefont {M.}~\bibnamefont
  {Collet}}, \bibinfo {author} {\bibfnamefont {O.}~\bibnamefont {Gladii}},
  \bibinfo {author} {\bibfnamefont {M.}~\bibnamefont {Evelt}}, \bibinfo
  {author} {\bibfnamefont {V.}~\bibnamefont {Bessonov}}, \bibinfo {author}
  {\bibfnamefont {L.}~\bibnamefont {Soumah}}, \bibinfo {author} {\bibfnamefont
  {P.}~\bibnamefont {Bortolotti}}, \bibinfo {author} {\bibfnamefont {S.~O.}\
  \bibnamefont {Demokritov}}, \bibinfo {author} {\bibfnamefont
  {Y.}~\bibnamefont {Henry}}, \bibinfo {author} {\bibfnamefont
  {V.}~\bibnamefont {Cros}}, \bibinfo {author} {\bibfnamefont {M.}~\bibnamefont
  {Bailleul}}, \bibinfo {author} {\bibfnamefont {V.~E.}\ \bibnamefont
  {Demidov}},\ and\ \bibinfo {author} {\bibfnamefont {A.}~\bibnamefont
  {Anane}},\ }\bibfield  {title} {\bibinfo {title} {Spin-wave propagation in
  ultra-thin yig based waveguides},\ }\href {https://doi.org/10.1063/1.4976708}
  {\bibfield  {journal} {\bibinfo  {journal} {Applied Physics Letters}\
  }\textbf {\bibinfo {volume} {110}},\ \bibinfo {pages} {092408} (\bibinfo
  {year} {2017}{\natexlab{a}})},\ \Eprint
  {https://arxiv.org/abs/https://doi.org/10.1063/1.4976708}
  {https://doi.org/10.1063/1.4976708} \BibitemShut {NoStop}%
\bibitem [{\citenamefont {Hansen}\ \emph {et~al.}(1974)\citenamefont {Hansen},
  \citenamefont {Röschmann},\ and\ \citenamefont {Tolksdorf}}]{Hansen1974}%
  \BibitemOpen
  \bibfield  {author} {\bibinfo {author} {\bibfnamefont {P.}~\bibnamefont
  {Hansen}}, \bibinfo {author} {\bibfnamefont {P.}~\bibnamefont {Röschmann}},\
  and\ \bibinfo {author} {\bibfnamefont {W.}~\bibnamefont {Tolksdorf}},\
  }\bibfield  {title} {\bibinfo {title} {Saturation magnetization of
  gallium‐substituted yttrium iron garnet},\ }\href
  {https://doi.org/10.1063/1.1663657} {\bibfield  {journal} {\bibinfo
  {journal} {Journal of Applied Physics}\ }\textbf {\bibinfo {volume} {45}},\
  \bibinfo {pages} {2728} (\bibinfo {year} {1974})},\ \Eprint
  {https://arxiv.org/abs/https://doi.org/10.1063/1.1663657}
  {https://doi.org/10.1063/1.1663657} \BibitemShut {NoStop}%
\bibitem [{\citenamefont {Kittel}(1948)}]{Kittel1948}%
  \BibitemOpen
  \bibfield  {author} {\bibinfo {author} {\bibfnamefont {C.}~\bibnamefont
  {Kittel}},\ }\bibfield  {title} {\bibinfo {title} {On the theory of
  ferromagnetic resonance absorption},\ }\href
  {https://doi.org/10.1103/PhysRev.73.155} {\bibfield  {journal} {\bibinfo
  {journal} {Phys. Rev.}\ }\textbf {\bibinfo {volume} {73}},\ \bibinfo {pages}
  {155} (\bibinfo {year} {1948})}\BibitemShut {NoStop}%
\bibitem [{\citenamefont {Manuilov}\ \emph {et~al.}(2009)\citenamefont
  {Manuilov}, \citenamefont {Fors}, \citenamefont {Khartsev},\ and\
  \citenamefont {Grishin}}]{Manuilov2009}%
  \BibitemOpen
  \bibfield  {author} {\bibinfo {author} {\bibfnamefont {S.~A.}\ \bibnamefont
  {Manuilov}}, \bibinfo {author} {\bibfnamefont {R.}~\bibnamefont {Fors}},
  \bibinfo {author} {\bibfnamefont {S.~I.}\ \bibnamefont {Khartsev}},\ and\
  \bibinfo {author} {\bibfnamefont {A.~M.}\ \bibnamefont {Grishin}},\
  }\bibfield  {title} {\bibinfo {title} {Submicron y3fe5o12 film magnetostatic
  wave band pass filters},\ }\href {https://doi.org/10.1063/1.3075816}
  {\bibfield  {journal} {\bibinfo  {journal} {Journal of Applied Physics}\
  }\textbf {\bibinfo {volume} {105}},\ \bibinfo {pages} {033917} (\bibinfo
  {year} {2009})},\ \Eprint
  {https://arxiv.org/abs/https://doi.org/10.1063/1.3075816}
  {https://doi.org/10.1063/1.3075816} \BibitemShut {NoStop}%
\bibitem [{\citenamefont {{McMichael}}\ and\ \citenamefont
  {{Krivosik}}(2004)}]{Mcmichael2004}%
  \BibitemOpen
  \bibfield  {author} {\bibinfo {author} {\bibfnamefont {R.~D.}\ \bibnamefont
  {{McMichael}}}\ and\ \bibinfo {author} {\bibfnamefont {P.}~\bibnamefont
  {{Krivosik}}},\ }\bibfield  {title} {\bibinfo {title} {{Classical Model of
  Extrinsic Ferromagnetic Resonance Linewidth in Ultrathin Films}},\ }\href
  {https://doi.org/10.1109/TMAG.2003.821564} {\bibfield  {journal} {\bibinfo
  {journal} {IEEE Transactions on Magnetics}\ }\textbf {\bibinfo {volume}
  {40}},\ \bibinfo {pages} {2} (\bibinfo {year} {2004})}\BibitemShut {NoStop}%
\bibitem [{\citenamefont {Sun}\ \emph {et~al.}(2012)\citenamefont {Sun},
  \citenamefont {Song}, \citenamefont {Chang}, \citenamefont {Kabatek},
  \citenamefont {Jantz}, \citenamefont {Schneider}, \citenamefont {Wu},
  \citenamefont {Schultheiss},\ and\ \citenamefont {Hoffmann}}]{Sun2012}%
  \BibitemOpen
  \bibfield  {author} {\bibinfo {author} {\bibfnamefont {Y.}~\bibnamefont
  {Sun}}, \bibinfo {author} {\bibfnamefont {Y.-Y.}\ \bibnamefont {Song}},
  \bibinfo {author} {\bibfnamefont {H.}~\bibnamefont {Chang}}, \bibinfo
  {author} {\bibfnamefont {M.}~\bibnamefont {Kabatek}}, \bibinfo {author}
  {\bibfnamefont {M.}~\bibnamefont {Jantz}}, \bibinfo {author} {\bibfnamefont
  {W.}~\bibnamefont {Schneider}}, \bibinfo {author} {\bibfnamefont
  {M.}~\bibnamefont {Wu}}, \bibinfo {author} {\bibfnamefont {H.}~\bibnamefont
  {Schultheiss}},\ and\ \bibinfo {author} {\bibfnamefont {A.}~\bibnamefont
  {Hoffmann}},\ }\bibfield  {title} {\bibinfo {title} {Growth and ferromagnetic
  resonance properties of nanometer-thick yttrium iron garnet films},\ }\href
  {https://doi.org/10.1063/1.4759039} {\bibfield  {journal} {\bibinfo
  {journal} {Applied Physics Letters}\ }\textbf {\bibinfo {volume} {101}},\
  \bibinfo {pages} {152405} (\bibinfo {year} {2012})},\ \Eprint
  {https://arxiv.org/abs/https://doi.org/10.1063/1.4759039}
  {https://doi.org/10.1063/1.4759039} \BibitemShut {NoStop}%
\bibitem [{\citenamefont {d'Allivy Kelly}\ \emph {et~al.}(2013)\citenamefont
  {d'Allivy Kelly}, \citenamefont {Anane}, \citenamefont {Bernard},
  \citenamefont {Ben~Youssef}, \citenamefont {Hahn}, \citenamefont
  {Molpeceres}, \citenamefont {Carrétéro}, \citenamefont {Jacquet},
  \citenamefont {Deranlot}, \citenamefont {Bortolotti}, \citenamefont
  {Lebourgeois}, \citenamefont {Mage}, \citenamefont {de~Loubens},
  \citenamefont {Klein}, \citenamefont {Cros},\ and\ \citenamefont
  {Fert}}]{Kelly2013}%
  \BibitemOpen
  \bibfield  {author} {\bibinfo {author} {\bibfnamefont {O.}~\bibnamefont
  {d'Allivy Kelly}}, \bibinfo {author} {\bibfnamefont {A.}~\bibnamefont
  {Anane}}, \bibinfo {author} {\bibfnamefont {R.}~\bibnamefont {Bernard}},
  \bibinfo {author} {\bibfnamefont {J.}~\bibnamefont {Ben~Youssef}}, \bibinfo
  {author} {\bibfnamefont {C.}~\bibnamefont {Hahn}}, \bibinfo {author}
  {\bibfnamefont {A.~H.}\ \bibnamefont {Molpeceres}}, \bibinfo {author}
  {\bibfnamefont {C.}~\bibnamefont {Carrétéro}}, \bibinfo {author}
  {\bibfnamefont {E.}~\bibnamefont {Jacquet}}, \bibinfo {author} {\bibfnamefont
  {C.}~\bibnamefont {Deranlot}}, \bibinfo {author} {\bibfnamefont
  {P.}~\bibnamefont {Bortolotti}}, \bibinfo {author} {\bibfnamefont
  {R.}~\bibnamefont {Lebourgeois}}, \bibinfo {author} {\bibfnamefont {J.-C.}\
  \bibnamefont {Mage}}, \bibinfo {author} {\bibfnamefont {G.}~\bibnamefont
  {de~Loubens}}, \bibinfo {author} {\bibfnamefont {O.}~\bibnamefont {Klein}},
  \bibinfo {author} {\bibfnamefont {V.}~\bibnamefont {Cros}},\ and\ \bibinfo
  {author} {\bibfnamefont {A.}~\bibnamefont {Fert}},\ }\bibfield  {title}
  {\bibinfo {title} {Inverse spin hall effect in nanometer-thick yttrium iron
  garnet/pt system},\ }\href {https://doi.org/10.1063/1.4819157} {\bibfield
  {journal} {\bibinfo  {journal} {Applied Physics Letters}\ }\textbf {\bibinfo
  {volume} {103}},\ \bibinfo {pages} {082408} (\bibinfo {year} {2013})},\
  \Eprint {https://arxiv.org/abs/https://doi.org/10.1063/1.4819157}
  {https://doi.org/10.1063/1.4819157} \BibitemShut {NoStop}%
\bibitem [{\citenamefont {Onbasli}\ \emph {et~al.}(2014)\citenamefont
  {Onbasli}, \citenamefont {Kehlberger}, \citenamefont {Kim}, \citenamefont
  {Jakob}, \citenamefont {Kläui}, \citenamefont {Chumak}, \citenamefont
  {Hillebrands},\ and\ \citenamefont {Ross}}]{Onbasli2014}%
  \BibitemOpen
  \bibfield  {author} {\bibinfo {author} {\bibfnamefont {M.~C.}\ \bibnamefont
  {Onbasli}}, \bibinfo {author} {\bibfnamefont {A.}~\bibnamefont {Kehlberger}},
  \bibinfo {author} {\bibfnamefont {D.~H.}\ \bibnamefont {Kim}}, \bibinfo
  {author} {\bibfnamefont {G.}~\bibnamefont {Jakob}}, \bibinfo {author}
  {\bibfnamefont {M.}~\bibnamefont {Kläui}}, \bibinfo {author} {\bibfnamefont
  {A.~V.}\ \bibnamefont {Chumak}}, \bibinfo {author} {\bibfnamefont
  {B.}~\bibnamefont {Hillebrands}},\ and\ \bibinfo {author} {\bibfnamefont
  {C.~A.}\ \bibnamefont {Ross}},\ }\bibfield  {title} {\bibinfo {title} {Pulsed
  laser deposition of epitaxial yttrium iron garnet films with low gilbert
  damping and bulk-like magnetization},\ }\href
  {https://doi.org/10.1063/1.4896936} {\bibfield  {journal} {\bibinfo
  {journal} {APL Materials}\ }\textbf {\bibinfo {volume} {2}},\ \bibinfo
  {pages} {106102} (\bibinfo {year} {2014})},\ \Eprint
  {https://arxiv.org/abs/https://doi.org/10.1063/1.4896936}
  {https://doi.org/10.1063/1.4896936} \BibitemShut {NoStop}%
\bibitem [{\citenamefont {Howe}\ \emph {et~al.}(2015)\citenamefont {Howe},
  \citenamefont {Emori}, \citenamefont {Jeon}, \citenamefont {Oxholm},
  \citenamefont {Jones}, \citenamefont {Mahalingam}, \citenamefont {Zhuang},
  \citenamefont {Sun},\ and\ \citenamefont {Brown}}]{Howe2015}%
  \BibitemOpen
  \bibfield  {author} {\bibinfo {author} {\bibfnamefont {B.~M.}\ \bibnamefont
  {Howe}}, \bibinfo {author} {\bibfnamefont {S.}~\bibnamefont {Emori}},
  \bibinfo {author} {\bibfnamefont {H.}~\bibnamefont {Jeon}}, \bibinfo {author}
  {\bibfnamefont {T.~M.}\ \bibnamefont {Oxholm}}, \bibinfo {author}
  {\bibfnamefont {J.~G.}\ \bibnamefont {Jones}}, \bibinfo {author}
  {\bibfnamefont {K.}~\bibnamefont {Mahalingam}}, \bibinfo {author}
  {\bibfnamefont {Y.}~\bibnamefont {Zhuang}}, \bibinfo {author} {\bibfnamefont
  {N.~X.}\ \bibnamefont {Sun}},\ and\ \bibinfo {author} {\bibfnamefont {G.~J.}\
  \bibnamefont {Brown}},\ }\bibfield  {title} {\bibinfo {title} {Pseudomorphic
  yttrium iron garnet thin films with low damping and inhomogeneous linewidth
  broadening},\ }\bibfield  {booktitle} {\emph {\bibinfo {booktitle} {IEEE
  Magnetics Letters}},\ }\href {https://doi.org/10.1109/LMAG.2015.2449260}
  {\bibfield  {journal} {\bibinfo  {journal} {IEEE Magnetics Letters}\ }\textbf
  {\bibinfo {volume} {6}},\ \bibinfo {pages} {1} (\bibinfo {year}
  {2015})}\BibitemShut {NoStop}%
\bibitem [{\citenamefont {Tang}\ \emph {et~al.}(2016)\citenamefont {Tang},
  \citenamefont {Aldosary}, \citenamefont {Jiang}, \citenamefont {Chang},
  \citenamefont {Madon}, \citenamefont {Chan}, \citenamefont {Wu},
  \citenamefont {Garay},\ and\ \citenamefont {Shi}}]{Tang2016}%
  \BibitemOpen
  \bibfield  {author} {\bibinfo {author} {\bibfnamefont {C.}~\bibnamefont
  {Tang}}, \bibinfo {author} {\bibfnamefont {M.}~\bibnamefont {Aldosary}},
  \bibinfo {author} {\bibfnamefont {Z.}~\bibnamefont {Jiang}}, \bibinfo
  {author} {\bibfnamefont {H.}~\bibnamefont {Chang}}, \bibinfo {author}
  {\bibfnamefont {B.}~\bibnamefont {Madon}}, \bibinfo {author} {\bibfnamefont
  {K.}~\bibnamefont {Chan}}, \bibinfo {author} {\bibfnamefont {M.}~\bibnamefont
  {Wu}}, \bibinfo {author} {\bibfnamefont {J.~E.}\ \bibnamefont {Garay}},\ and\
  \bibinfo {author} {\bibfnamefont {J.}~\bibnamefont {Shi}},\ }\bibfield
  {title} {\bibinfo {title} {Exquisite growth control and magnetic properties
  of yttrium iron garnet thin films},\ }\href
  {https://doi.org/10.1063/1.4943210} {\bibfield  {journal} {\bibinfo
  {journal} {Applied Physics Letters}\ }\textbf {\bibinfo {volume} {108}},\
  \bibinfo {pages} {102403} (\bibinfo {year} {2016})},\ \Eprint
  {https://arxiv.org/abs/https://doi.org/10.1063/1.4943210}
  {https://doi.org/10.1063/1.4943210} \BibitemShut {NoStop}%
\bibitem [{\citenamefont {Hahn}\ \emph {et~al.}(2014)\citenamefont {Hahn},
  \citenamefont {Naletov}, \citenamefont {de~Loubens}, \citenamefont {Klein},
  \citenamefont {d'Allivy Kelly}, \citenamefont {Anane}, \citenamefont
  {Bernard}, \citenamefont {Jacquet}, \citenamefont {Bortolotti}, \citenamefont
  {Cros}, \citenamefont {Prieto},\ and\ \citenamefont {Muñoz}}]{Hahn2014}%
  \BibitemOpen
  \bibfield  {author} {\bibinfo {author} {\bibfnamefont {C.}~\bibnamefont
  {Hahn}}, \bibinfo {author} {\bibfnamefont {V.~V.}\ \bibnamefont {Naletov}},
  \bibinfo {author} {\bibfnamefont {G.}~\bibnamefont {de~Loubens}}, \bibinfo
  {author} {\bibfnamefont {O.}~\bibnamefont {Klein}}, \bibinfo {author}
  {\bibfnamefont {O.}~\bibnamefont {d'Allivy Kelly}}, \bibinfo {author}
  {\bibfnamefont {A.}~\bibnamefont {Anane}}, \bibinfo {author} {\bibfnamefont
  {R.}~\bibnamefont {Bernard}}, \bibinfo {author} {\bibfnamefont
  {E.}~\bibnamefont {Jacquet}}, \bibinfo {author} {\bibfnamefont
  {P.}~\bibnamefont {Bortolotti}}, \bibinfo {author} {\bibfnamefont
  {V.}~\bibnamefont {Cros}}, \bibinfo {author} {\bibfnamefont {J.~L.}\
  \bibnamefont {Prieto}},\ and\ \bibinfo {author} {\bibfnamefont
  {M.}~\bibnamefont {Muñoz}},\ }\bibfield  {title} {\bibinfo {title}
  {Measurement of the intrinsic damping constant in individual nanodisks of
  y3fe5o12 and y3fe5o12|pt},\ }\href {https://doi.org/10.1063/1.4871516}
  {\bibfield  {journal} {\bibinfo  {journal} {Applied Physics Letters}\
  }\textbf {\bibinfo {volume} {104}},\ \bibinfo {pages} {152410} (\bibinfo
  {year} {2014})},\ \Eprint
  {https://arxiv.org/abs/https://doi.org/10.1063/1.4871516}
  {https://doi.org/10.1063/1.4871516} \BibitemShut {NoStop}%
\bibitem [{\citenamefont {Collet}\ \emph
  {et~al.}(2017{\natexlab{b}})\citenamefont {Collet}, \citenamefont {Soumah},
  \citenamefont {Bortolotti}, \citenamefont {Muñoz}, \citenamefont {Cros},\
  and\ \citenamefont {Anane}}]{Collet2017_AIP}%
  \BibitemOpen
  \bibfield  {author} {\bibinfo {author} {\bibfnamefont {M.}~\bibnamefont
  {Collet}}, \bibinfo {author} {\bibfnamefont {L.}~\bibnamefont {Soumah}},
  \bibinfo {author} {\bibfnamefont {P.}~\bibnamefont {Bortolotti}}, \bibinfo
  {author} {\bibfnamefont {M.}~\bibnamefont {Muñoz}}, \bibinfo {author}
  {\bibfnamefont {V.}~\bibnamefont {Cros}},\ and\ \bibinfo {author}
  {\bibfnamefont {A.}~\bibnamefont {Anane}},\ }\bibfield  {title} {\bibinfo
  {title} {Spin seebeck effect in nanometer-thick yig micro-fabricated
  strips},\ }\href {https://doi.org/10.1063/1.4976332} {\bibfield  {journal}
  {\bibinfo  {journal} {AIP Advances}\ }\textbf {\bibinfo {volume} {7}},\
  \bibinfo {pages} {055924} (\bibinfo {year} {2017}{\natexlab{b}})},\ \Eprint
  {https://arxiv.org/abs/https://doi.org/10.1063/1.4976332}
  {https://doi.org/10.1063/1.4976332} \BibitemShut {NoStop}%
\bibitem [{\citenamefont {Jungfleisch}\ \emph {et~al.}(2015)\citenamefont
  {Jungfleisch}, \citenamefont {Zhang}, \citenamefont {Jiang}, \citenamefont
  {Chang}, \citenamefont {Sklenar}, \citenamefont {Wu}, \citenamefont
  {Pearson}, \citenamefont {Bhattacharya}, \citenamefont {Ketterson},
  \citenamefont {Wu},\ and\ \citenamefont {Hoffmann}}]{Jungfleisch2015}%
  \BibitemOpen
  \bibfield  {author} {\bibinfo {author} {\bibfnamefont {M.~B.}\ \bibnamefont
  {Jungfleisch}}, \bibinfo {author} {\bibfnamefont {W.}~\bibnamefont {Zhang}},
  \bibinfo {author} {\bibfnamefont {W.}~\bibnamefont {Jiang}}, \bibinfo
  {author} {\bibfnamefont {H.}~\bibnamefont {Chang}}, \bibinfo {author}
  {\bibfnamefont {J.}~\bibnamefont {Sklenar}}, \bibinfo {author} {\bibfnamefont
  {S.~M.}\ \bibnamefont {Wu}}, \bibinfo {author} {\bibfnamefont {J.~E.}\
  \bibnamefont {Pearson}}, \bibinfo {author} {\bibfnamefont {A.}~\bibnamefont
  {Bhattacharya}}, \bibinfo {author} {\bibfnamefont {J.~B.}\ \bibnamefont
  {Ketterson}}, \bibinfo {author} {\bibfnamefont {M.}~\bibnamefont {Wu}},\ and\
  \bibinfo {author} {\bibfnamefont {A.}~\bibnamefont {Hoffmann}},\ }\bibfield
  {title} {\bibinfo {title} {Spin waves in micro-structured yttrium iron garnet
  nanometer-thick films},\ }\href {https://doi.org/10.1063/1.4916027}
  {\bibfield  {journal} {\bibinfo  {journal} {Journal of Applied Physics}\
  }\textbf {\bibinfo {volume} {117}},\ \bibinfo {pages} {17D128} (\bibinfo
  {year} {2015})},\ \Eprint
  {https://arxiv.org/abs/https://doi.org/10.1063/1.4916027}
  {https://doi.org/10.1063/1.4916027} \BibitemShut {NoStop}%
\bibitem [{\citenamefont {{Yang}}\ and\ \citenamefont
  {{Hammel}}(2018)}]{Yang2018}%
  \BibitemOpen
  \bibfield  {author} {\bibinfo {author} {\bibfnamefont {F.}~\bibnamefont
  {{Yang}}}\ and\ \bibinfo {author} {\bibfnamefont {P.~C.}\ \bibnamefont
  {{Hammel}}},\ }\bibfield  {title} {\bibinfo {title} {{FMR-driven spin pumping
  in Y$_{3}$Fe$_{5}$O$_{12}$-based structures}},\ }\href
  {https://doi.org/10.1088/1361-6463/aac249} {\bibfield  {journal} {\bibinfo
  {journal} {Journal of Physics D Applied Physics}\ }\textbf {\bibinfo {volume}
  {51}},\ \bibinfo {eid} {253001} (\bibinfo {year} {2018})}\BibitemShut
  {NoStop}%
\bibitem [{\citenamefont {{Wang}}\ \emph {et~al.}(2013)\citenamefont {{Wang}},
  \citenamefont {{Du}}, \citenamefont {{Pu}}, \citenamefont {{Adur}},
  \citenamefont {{Hammel}},\ and\ \citenamefont {{Yang}}}]{Wang2013}%
  \BibitemOpen
  \bibfield  {author} {\bibinfo {author} {\bibfnamefont {H.~L.}\ \bibnamefont
  {{Wang}}}, \bibinfo {author} {\bibfnamefont {C.~H.}\ \bibnamefont {{Du}}},
  \bibinfo {author} {\bibfnamefont {Y.}~\bibnamefont {{Pu}}}, \bibinfo {author}
  {\bibfnamefont {R.}~\bibnamefont {{Adur}}}, \bibinfo {author} {\bibfnamefont
  {P.~C.}\ \bibnamefont {{Hammel}}},\ and\ \bibinfo {author} {\bibfnamefont
  {F.~Y.}\ \bibnamefont {{Yang}}},\ }\bibfield  {title} {\bibinfo {title}
  {{Large spin pumping from epitaxial Y$_{3}$Fe$_{5}$O$_{12}$ thin films to Pt
  and W layers}},\ }\href {https://doi.org/10.1103/PhysRevB.88.100406}
  {\bibfield  {journal} {\bibinfo  {journal} {Physical Review B}\ }\textbf
  {\bibinfo {volume} {88}},\ \bibinfo {eid} {100406} (\bibinfo {year}
  {2013})},\ \Eprint {https://arxiv.org/abs/1307.1172} {arXiv:1307.1172
  [cond-mat.mtrl-sci]} \BibitemShut {NoStop}%
\bibitem [{\citenamefont {Chang}\ \emph {et~al.}(2017)\citenamefont {Chang},
  \citenamefont {Praveen~Janantha}, \citenamefont {Ding}, \citenamefont {Liu},
  \citenamefont {Cline}, \citenamefont {Gelfand}, \citenamefont {Li},
  \citenamefont {Marconi},\ and\ \citenamefont {Wu}}]{Chang2017}%
  \BibitemOpen
  \bibfield  {author} {\bibinfo {author} {\bibfnamefont {H.}~\bibnamefont
  {Chang}}, \bibinfo {author} {\bibfnamefont {P.~A.}\ \bibnamefont
  {Praveen~Janantha}}, \bibinfo {author} {\bibfnamefont {J.}~\bibnamefont
  {Ding}}, \bibinfo {author} {\bibfnamefont {T.}~\bibnamefont {Liu}}, \bibinfo
  {author} {\bibfnamefont {K.}~\bibnamefont {Cline}}, \bibinfo {author}
  {\bibfnamefont {J.~N.}\ \bibnamefont {Gelfand}}, \bibinfo {author}
  {\bibfnamefont {W.}~\bibnamefont {Li}}, \bibinfo {author} {\bibfnamefont
  {M.~C.}\ \bibnamefont {Marconi}},\ and\ \bibinfo {author} {\bibfnamefont
  {M.}~\bibnamefont {Wu}},\ }\bibfield  {title} {\bibinfo {title} {Role of
  damping in spin seebeck effect in yttrium iron garnet thin films},\
  }\bibfield  {journal} {\bibinfo  {journal} {Science Advances}\ }\textbf
  {\bibinfo {volume} {3}},\ \href {https://doi.org/10.1126/sciadv.1601614}
  {10.1126/sciadv.1601614} (\bibinfo {year} {2017}),\ \Eprint
  {https://arxiv.org/abs/https://advances.sciencemag.org/content/3/4/e1601614}
  {https://advances.sciencemag.org/content/3/4/e1601614} \BibitemShut {NoStop}%
\bibitem [{\citenamefont {Lustikova}\ \emph {et~al.}(2014)\citenamefont
  {Lustikova}, \citenamefont {Shiomi}, \citenamefont {Qiu}, \citenamefont
  {Kikkawa}, \citenamefont {Iguchi}, \citenamefont {Uchida},\ and\
  \citenamefont {Saitoh}}]{Lustikova2014}%
  \BibitemOpen
  \bibfield  {author} {\bibinfo {author} {\bibfnamefont {J.}~\bibnamefont
  {Lustikova}}, \bibinfo {author} {\bibfnamefont {Y.}~\bibnamefont {Shiomi}},
  \bibinfo {author} {\bibfnamefont {Z.}~\bibnamefont {Qiu}}, \bibinfo {author}
  {\bibfnamefont {T.}~\bibnamefont {Kikkawa}}, \bibinfo {author} {\bibfnamefont
  {R.}~\bibnamefont {Iguchi}}, \bibinfo {author} {\bibfnamefont
  {K.}~\bibnamefont {Uchida}},\ and\ \bibinfo {author} {\bibfnamefont
  {E.}~\bibnamefont {Saitoh}},\ }\bibfield  {title} {\bibinfo {title} {Spin
  current generation from sputtered y3fe5o12 films},\ }\href
  {https://doi.org/10.1063/1.4898161} {\bibfield  {journal} {\bibinfo
  {journal} {Journal of Applied Physics}\ }\textbf {\bibinfo {volume} {116}},\
  \bibinfo {pages} {153902} (\bibinfo {year} {2014})},\ \Eprint
  {https://arxiv.org/abs/https://doi.org/10.1063/1.4898161}
  {https://doi.org/10.1063/1.4898161} \BibitemShut {NoStop}%
\bibitem [{\citenamefont {Hauser}\ \emph {et~al.}(2017)\citenamefont {Hauser},
  \citenamefont {Eisenschmidt}, \citenamefont {Richter}, \citenamefont
  {Müller}, \citenamefont {Deniz},\ and\ \citenamefont
  {Schmidt}}]{Hauser2017}%
  \BibitemOpen
  \bibfield  {author} {\bibinfo {author} {\bibfnamefont {C.}~\bibnamefont
  {Hauser}}, \bibinfo {author} {\bibfnamefont {C.}~\bibnamefont
  {Eisenschmidt}}, \bibinfo {author} {\bibfnamefont {T.}~\bibnamefont
  {Richter}}, \bibinfo {author} {\bibfnamefont {A.}~\bibnamefont {Müller}},
  \bibinfo {author} {\bibfnamefont {H.}~\bibnamefont {Deniz}},\ and\ \bibinfo
  {author} {\bibfnamefont {G.}~\bibnamefont {Schmidt}},\ }\bibfield  {title}
  {\bibinfo {title} {Annealing of amorphous yttrium iron garnet thin films in
  argon atmosphere},\ }\href {https://doi.org/10.1063/1.4999829} {\bibfield
  {journal} {\bibinfo  {journal} {Journal of Applied Physics}\ }\textbf
  {\bibinfo {volume} {122}},\ \bibinfo {pages} {083908} (\bibinfo {year}
  {2017})},\ \Eprint {https://arxiv.org/abs/https://doi.org/10.1063/1.4999829}
  {https://doi.org/10.1063/1.4999829} \BibitemShut {NoStop}%
\bibitem [{\citenamefont {Krysztofik}\ \emph {et~al.}(2017)\citenamefont
  {Krysztofik}, \citenamefont {Coy}, \citenamefont {Kuswik}, \citenamefont
  {Zaleski}, \citenamefont {Glowinski},\ and\ \citenamefont
  {Dubowik}}]{Krysztofik2017}%
  \BibitemOpen
  \bibfield  {author} {\bibinfo {author} {\bibfnamefont {A.}~\bibnamefont
  {Krysztofik}}, \bibinfo {author} {\bibfnamefont {L.~E.}\ \bibnamefont {Coy}},
  \bibinfo {author} {\bibfnamefont {P.}~\bibnamefont {Kuswik}}, \bibinfo
  {author} {\bibfnamefont {K.}~\bibnamefont {Zaleski}}, \bibinfo {author}
  {\bibfnamefont {H.}~\bibnamefont {Glowinski}},\ and\ \bibinfo {author}
  {\bibfnamefont {J.}~\bibnamefont {Dubowik}},\ }\bibfield  {title} {\bibinfo
  {title} {Ultra-low damping in lift-off structured yttrium iron garnet thin
  films},\ }\href {https://doi.org/10.1063/1.5002004} {\bibfield  {journal}
  {\bibinfo  {journal} {Applied Physics Letters}\ }\textbf {\bibinfo {volume}
  {111}},\ \bibinfo {pages} {192404} (\bibinfo {year} {2017})},\ \Eprint
  {https://arxiv.org/abs/https://doi.org/10.1063/1.5002004}
  {https://doi.org/10.1063/1.5002004} \BibitemShut {NoStop}%
\bibitem [{\citenamefont {Li}\ \emph {et~al.}(2016)\citenamefont {Li},
  \citenamefont {Zhang}, \citenamefont {Ding}, \citenamefont {Pearson},
  \citenamefont {Novosad},\ and\ \citenamefont {Hoffmann}}]{Li2016}%
  \BibitemOpen
  \bibfield  {author} {\bibinfo {author} {\bibfnamefont {S.}~\bibnamefont
  {Li}}, \bibinfo {author} {\bibfnamefont {W.}~\bibnamefont {Zhang}}, \bibinfo
  {author} {\bibfnamefont {J.}~\bibnamefont {Ding}}, \bibinfo {author}
  {\bibfnamefont {J.~E.}\ \bibnamefont {Pearson}}, \bibinfo {author}
  {\bibfnamefont {V.}~\bibnamefont {Novosad}},\ and\ \bibinfo {author}
  {\bibfnamefont {A.}~\bibnamefont {Hoffmann}},\ }\bibfield  {title} {\bibinfo
  {title} {Epitaxial patterning of nanometer-thick y3fe5o12 films with low
  magnetic damping},\ }\href {https://doi.org/10.1039/C5NR06808H} {\bibfield
  {journal} {\bibinfo  {journal} {Nanoscale}\ }\textbf {\bibinfo {volume}
  {8}},\ \bibinfo {pages} {388} (\bibinfo {year} {2016})}\BibitemShut {NoStop}%
\bibitem [{\citenamefont {Jungfleisch}\ \emph {et~al.}(2017)\citenamefont
  {Jungfleisch}, \citenamefont {Ding}, \citenamefont {Zhang}, \citenamefont
  {Jiang}, \citenamefont {Pearson}, \citenamefont {Novosad},\ and\
  \citenamefont {Hoffmann}}]{Jungfleisch2017}%
  \BibitemOpen
  \bibfield  {author} {\bibinfo {author} {\bibfnamefont {M.~B.}\ \bibnamefont
  {Jungfleisch}}, \bibinfo {author} {\bibfnamefont {J.}~\bibnamefont {Ding}},
  \bibinfo {author} {\bibfnamefont {W.}~\bibnamefont {Zhang}}, \bibinfo
  {author} {\bibfnamefont {W.}~\bibnamefont {Jiang}}, \bibinfo {author}
  {\bibfnamefont {J.~E.}\ \bibnamefont {Pearson}}, \bibinfo {author}
  {\bibfnamefont {V.}~\bibnamefont {Novosad}},\ and\ \bibinfo {author}
  {\bibfnamefont {A.}~\bibnamefont {Hoffmann}},\ }\bibfield  {title} {\bibinfo
  {title} {Insulating nanomagnets driven by spin torque},\ }\bibfield
  {booktitle} {\emph {\bibinfo {booktitle} {Nano Letters}},\ }\href
  {https://doi.org/10.1021/acs.nanolett.6b02794} {\bibfield  {journal}
  {\bibinfo  {journal} {Nano Letters}\ }\textbf {\bibinfo {volume} {17}},\
  \bibinfo {pages} {8} (\bibinfo {year} {2017})}\BibitemShut {NoStop}%
\bibitem [{\citenamefont {Zhu}\ \emph {et~al.}(2017)\citenamefont {Zhu},
  \citenamefont {Chang}, \citenamefont {Franson}, \citenamefont {Liu},
  \citenamefont {Zhang}, \citenamefont {Johnston-Halperin}, \citenamefont
  {Wu},\ and\ \citenamefont {Tang}}]{Zhu2017}%
  \BibitemOpen
  \bibfield  {author} {\bibinfo {author} {\bibfnamefont {N.}~\bibnamefont
  {Zhu}}, \bibinfo {author} {\bibfnamefont {H.}~\bibnamefont {Chang}}, \bibinfo
  {author} {\bibfnamefont {A.}~\bibnamefont {Franson}}, \bibinfo {author}
  {\bibfnamefont {T.}~\bibnamefont {Liu}}, \bibinfo {author} {\bibfnamefont
  {X.}~\bibnamefont {Zhang}}, \bibinfo {author} {\bibfnamefont
  {E.}~\bibnamefont {Johnston-Halperin}}, \bibinfo {author} {\bibfnamefont
  {M.}~\bibnamefont {Wu}},\ and\ \bibinfo {author} {\bibfnamefont {H.~X.}\
  \bibnamefont {Tang}},\ }\bibfield  {title} {\bibinfo {title} {Patterned
  growth of crystalline y3fe5o12 nanostructures with engineered magnetic shape
  anisotropy},\ }\href {https://doi.org/10.1063/1.4986474} {\bibfield
  {journal} {\bibinfo  {journal} {Applied Physics Letters}\ }\textbf {\bibinfo
  {volume} {110}},\ \bibinfo {pages} {252401} (\bibinfo {year} {2017})},\
  \Eprint {https://arxiv.org/abs/https://doi.org/10.1063/1.4986474}
  {https://doi.org/10.1063/1.4986474} \BibitemShut {NoStop}%
\bibitem [{\citenamefont {{Mitra}}\ \emph {et~al.}(2017)\citenamefont
  {{Mitra}}, \citenamefont {{Cespedes}}, \citenamefont {{Ramasse}},
  \citenamefont {{Ali}}, \citenamefont {{Marmion}}, \citenamefont {{Ward}},
  \citenamefont {{Brydson}}, \citenamefont {{Kinane}}, \citenamefont
  {{Cooper}}, \citenamefont {{Langridge}},\ and\ \citenamefont
  {{Hickey}}}]{Mitra2017}%
  \BibitemOpen
  \bibfield  {author} {\bibinfo {author} {\bibfnamefont {A.}~\bibnamefont
  {{Mitra}}}, \bibinfo {author} {\bibfnamefont {O.}~\bibnamefont {{Cespedes}}},
  \bibinfo {author} {\bibfnamefont {Q.}~\bibnamefont {{Ramasse}}}, \bibinfo
  {author} {\bibfnamefont {M.}~\bibnamefont {{Ali}}}, \bibinfo {author}
  {\bibfnamefont {S.}~\bibnamefont {{Marmion}}}, \bibinfo {author}
  {\bibfnamefont {M.}~\bibnamefont {{Ward}}}, \bibinfo {author} {\bibfnamefont
  {R.~M.~D.}\ \bibnamefont {{Brydson}}}, \bibinfo {author} {\bibfnamefont
  {C.~J.}\ \bibnamefont {{Kinane}}}, \bibinfo {author} {\bibfnamefont
  {J.~F.~K.}\ \bibnamefont {{Cooper}}}, \bibinfo {author} {\bibfnamefont
  {S.}~\bibnamefont {{Langridge}}},\ and\ \bibinfo {author} {\bibfnamefont
  {B.~J.}\ \bibnamefont {{Hickey}}},\ }\bibfield  {title} {\bibinfo {title}
  {{Interfacial Origin of the Magnetisation Suppression of Thin Film Yttrium
  Iron Garnet}},\ }\href {https://doi.org/10.1038/s41598-017-10281-6}
  {\bibfield  {journal} {\bibinfo  {journal} {Scientific Reports}\ }\textbf
  {\bibinfo {volume} {7}},\ \bibinfo {eid} {11774} (\bibinfo {year}
  {2017})}\BibitemShut {NoStop}%
\bibitem [{\citenamefont {{Cooper}}\ \emph {et~al.}(2017)\citenamefont
  {{Cooper}}, \citenamefont {{Kinane}}, \citenamefont {{Langridge}},
  \citenamefont {{Ali}}, \citenamefont {{Hickey}}, \citenamefont {{Niizeki}},
  \citenamefont {{Uchida}}, \citenamefont {{Saitoh}}, \citenamefont
  {{Ambaye}},\ and\ \citenamefont {{Glavic}}}]{Cooper2017}%
  \BibitemOpen
  \bibfield  {author} {\bibinfo {author} {\bibfnamefont {J.~F.~K.}\
  \bibnamefont {{Cooper}}}, \bibinfo {author} {\bibfnamefont {C.~J.}\
  \bibnamefont {{Kinane}}}, \bibinfo {author} {\bibfnamefont {S.}~\bibnamefont
  {{Langridge}}}, \bibinfo {author} {\bibfnamefont {M.}~\bibnamefont {{Ali}}},
  \bibinfo {author} {\bibfnamefont {B.~J.}\ \bibnamefont {{Hickey}}}, \bibinfo
  {author} {\bibfnamefont {T.}~\bibnamefont {{Niizeki}}}, \bibinfo {author}
  {\bibfnamefont {K.}~\bibnamefont {{Uchida}}}, \bibinfo {author}
  {\bibfnamefont {E.}~\bibnamefont {{Saitoh}}}, \bibinfo {author}
  {\bibfnamefont {H.}~\bibnamefont {{Ambaye}}},\ and\ \bibinfo {author}
  {\bibfnamefont {A.}~\bibnamefont {{Glavic}}},\ }\bibfield  {title} {\bibinfo
  {title} {{Unexpected structural and magnetic depth dependence of YIG thin
  films}},\ }\href {https://doi.org/10.1103/PhysRevB.96.104404} {\bibfield
  {journal} {\bibinfo  {journal} {Physical Review B}\ }\textbf {\bibinfo
  {volume} {96}},\ \bibinfo {eid} {104404} (\bibinfo {year} {2017})},\ \Eprint
  {https://arxiv.org/abs/1703.08752} {arXiv:1703.08752 [cond-mat.mtrl-sci]}
  \BibitemShut {NoStop}%
\bibitem [{\citenamefont {{Bai}}\ \emph {et~al.}(2019)\citenamefont {{Bai}},
  \citenamefont {{Zhan}}, \citenamefont {{Li}}, \citenamefont {{Su}},
  \citenamefont {{Zhu}}, \citenamefont {{Zhang}}, \citenamefont {{Zhu}},\ and\
  \citenamefont {{Cai}}}]{Bai2019}%
  \BibitemOpen
  \bibfield  {author} {\bibinfo {author} {\bibfnamefont {H.}~\bibnamefont
  {{Bai}}}, \bibinfo {author} {\bibfnamefont {X.~Z.}\ \bibnamefont {{Zhan}}},
  \bibinfo {author} {\bibfnamefont {G.}~\bibnamefont {{Li}}}, \bibinfo {author}
  {\bibfnamefont {J.}~\bibnamefont {{Su}}}, \bibinfo {author} {\bibfnamefont
  {Z.~Z.}\ \bibnamefont {{Zhu}}}, \bibinfo {author} {\bibfnamefont
  {Y.}~\bibnamefont {{Zhang}}}, \bibinfo {author} {\bibfnamefont
  {T.}~\bibnamefont {{Zhu}}},\ and\ \bibinfo {author} {\bibfnamefont {J.~W.}\
  \bibnamefont {{Cai}}},\ }\bibfield  {title} {\bibinfo {title}
  {{Characterization of YIG thin films and vacuum annealing effect by polarized
  neutron reflectometry and magnetotransport measurements}},\ }\href
  {https://doi.org/10.1063/1.5124832} {\bibfield  {journal} {\bibinfo
  {journal} {Applied Physics Letters}\ }\textbf {\bibinfo {volume} {115}},\
  \bibinfo {eid} {182401} (\bibinfo {year} {2019})}\BibitemShut {NoStop}%
\bibitem [{\citenamefont {{Suturin}}\ \emph {et~al.}(2018)\citenamefont
  {{Suturin}}, \citenamefont {{Korovin}}, \citenamefont {{Bursian}},
  \citenamefont {{Lutsev}}, \citenamefont {{Bourobina}}, \citenamefont
  {{Yakovlev}}, \citenamefont {{Montecchi}}, \citenamefont {{Pasquali}},
  \citenamefont {{Ukleev}}, \citenamefont {{Vorobiev}}, \citenamefont
  {{Devishvili}},\ and\ \citenamefont {{Sokolov}}}]{Suturin2018}%
  \BibitemOpen
  \bibfield  {author} {\bibinfo {author} {\bibfnamefont {S.~M.}\ \bibnamefont
  {{Suturin}}}, \bibinfo {author} {\bibfnamefont {A.~M.}\ \bibnamefont
  {{Korovin}}}, \bibinfo {author} {\bibfnamefont {V.~E.}\ \bibnamefont
  {{Bursian}}}, \bibinfo {author} {\bibfnamefont {L.~V.}\ \bibnamefont
  {{Lutsev}}}, \bibinfo {author} {\bibfnamefont {V.}~\bibnamefont
  {{Bourobina}}}, \bibinfo {author} {\bibfnamefont {N.~L.}\ \bibnamefont
  {{Yakovlev}}}, \bibinfo {author} {\bibfnamefont {M.}~\bibnamefont
  {{Montecchi}}}, \bibinfo {author} {\bibfnamefont {L.}~\bibnamefont
  {{Pasquali}}}, \bibinfo {author} {\bibfnamefont {V.}~\bibnamefont
  {{Ukleev}}}, \bibinfo {author} {\bibfnamefont {A.}~\bibnamefont
  {{Vorobiev}}}, \bibinfo {author} {\bibfnamefont {A.}~\bibnamefont
  {{Devishvili}}},\ and\ \bibinfo {author} {\bibfnamefont {N.~S.}\ \bibnamefont
  {{Sokolov}}},\ }\bibfield  {title} {\bibinfo {title} {{Role of gallium
  diffusion in the formation of a magnetically dead layer at the
  Y$_{3}$Fe$_{5}$O$_{12}$/Gd$_{3}$Ga$_{5}$O$_{12}$ epitaxial interface}},\
  }\href {https://doi.org/10.1103/PhysRevMaterials.2.104404} {\bibfield
  {journal} {\bibinfo  {journal} {Physical Review Materials}\ }\textbf
  {\bibinfo {volume} {2}},\ \bibinfo {eid} {104404} (\bibinfo {year} {2018})},\
  \Eprint {https://arxiv.org/abs/1811.01321} {arXiv:1811.01321
  [cond-mat.mtrl-sci]} \BibitemShut {NoStop}%
\end{thebibliography}
\end{document}